\documentstyle[psfig,emulateapj]{article}
\lefthead{Goldberg \& Leonard}
\righthead{Measuring Flexion}

\begin{document}

\title{Measuring Flexion}

\author{David M. Goldberg and Adrienne Leonard}
\affil{Department of Physics, Drexel University, Philadelphia, PA
  19104\\
(goldberg@drexel.edu, adrienne.leonard@drexel.edu)}

\begin{abstract}
We describe practical approaches to measuring flexion in observed
galaxies.  In particular, we look at the issues involved in using the
Shapelets and HOLICs techniques as means of extracting 2nd order
lensing information.  We also develop an extension of HOLICs to
estimate flexion in the presence of noise, and with a nearly isotropic
PSF.  We test both approaches in simple simulated lenses as well as a
sample of possible background sources from ACS observations of A1689.
We find that because noise is weighted differently in shapelets and
HOLICs approaches, that the correlation between measurements of the
same object is somewhat diminished, but produce similar scatter due to
measurement noise.
\end{abstract}
\keywords{
cosmology:observations -- galaxies:clusters:general --
galaxies:photometry -- gravitational lensing
}

\section{Introduction}

\subsection{Motivation}

Flexion has recently been introduced as a means of measuring small
scale variations in weak gravitational lens fields (Goldberg \& Bacon,
2005; Bacon, Goldberg, Rowe \& Taylor, 2006, hereafter BGRT).  Rather
than simply measuring the ellipticities of arclets, this technique
aims to measure the ``arciness'' and ``skewness'' (collectively
referred to as ``flexion'') of a lensed image.  Flexion is a
complementary approach to shear analysis in that it uses the odd
moments (3rd multipole moments, for example) to compute local
gradients in a shear field.  BGRT have discussed how flexion may be
used to identify substructure in clusters, to normalize the matter
power spectrum on sub-arcminute scales via ``cosmic flexion'' (as an
analog to cosmic shear), and to estimate the ellipticity of
galaxy-galaxy lenses.  As a practical application, flexion has already
been used to measure galaxy-galaxy lensing (Goldberg \& Bacon, 2005),
and is presently being used in cluster reconstruction (Leonard et al.,
in preparation).

However, there have been several difficulties in the estimation of
flexion on real objects.  First, the flexion inversion is difficult to
describe, contains an enormous number of terms, and thus, is rather
daunting to code.  Secondly, there has been little discussion of the
explicit effects of PSF convolution or deconvolution.  Finally, unlike
shear, there has, until recently, been no simple form to even
approximate what the ``flexion'' is.

The remainder of this paper will thus be a practical guide to
measuring flexion in real images.  We begin, below, by reminding the
reader of the basic terms involved in flexion analysis.  In
\S~\ref{sec:shapelets}, we review shapelet decomposition, and discuss
some of the issues involved in using shapelets to measure flexion.  In
\S~\ref{sec:holics}, we discuss a new, conceptually simpler, form of
flexion analysis developed by Okura et al. (2006), which uses moments,
rather than basis functions to measure flexion.  They call their
technique Higher Order Lensing Image's Characteristics, or HOLICs.  We
refine the HOLICs approach somewhat, and develop a KSB (Kaiser,
Squires, \& Broadhurst, 1995)-type approach using a Gaussian filter to
perform an inversion, as well as describe a technique for PSF
deconvolution.  In \S~\ref{sec:simulate}, we discuss comparisons of
the two techniques using simulated lenses and simulated PSFs.  In
\S~\ref{sec:measurement}, we compare shapelets and HOLICs inversions
on HST images.  Finally, in \S~\ref{sec:discuss}, we discuss the
implications of this study.

In Appendix A, we also present the explicit HOLICs inversion matrix,
so the reader can write his/her own code.  He/she need not do so,
however, as all codes discussed herein are available from the flexion
webpage.\footnote{http://www.physics.drexel.edu/\~{}goldberg/flexion}

\subsection{Flexion}

What is flexion?  Conceptually, flexion represents local variability
in the shear field which expresses itself as second-order distortions
in the coordinate transformation between unlensed and lensed images:
\begin{equation}
\beta_i\simeq A_{ij}\theta_j +\frac{1}{2}D_{ijk} \theta_j \theta_k,
\label{eq:transform}
\end{equation}
with
\begin{equation}
D_{ijk}=\partial_k A_{ij} \ ,
\end{equation}
where $\partial_k$ is shorthand for $\partial/\partial x_k$.
Here, ${\bf A}$ is the normal deprojection operator:
\begin{equation}
{\bf A}=
\left(
\begin{array}{cc}
1-\kappa-\gamma_1 & -\gamma_2 \\
-\gamma_2 & 1-\kappa+\gamma_1
\end{array}
\right)\ ,
\label{eq:Adef}
\end{equation}
and thus, the second term on right in equation~(\ref{eq:transform})
represents the flexion signal.  ${\bf D}$ may be written as:
\begin{eqnarray}
D_{ij1}&=&\left(
\begin{array}{cc}
-2\gamma_{1,1}-\gamma_{2,2} & \ \ -\gamma_{2,1} \\
-\gamma_{2,1} & \ \ -\gamma_{2,2}
\end{array}
\right), \\ \nonumber
D_{ij2}&=& \left(
\begin{array}{cc}
-\gamma_{2,1} & \ \ -\gamma_{2,2}\\
-\gamma_{2,2} & \ \ 2\gamma_{1,2}-\gamma_{2,1}
\end{array}
\right).
\end{eqnarray}

These distortions create asymmetries in a lensed image -- a skewness
and a bending, depending on the values of individual coefficients.
Irwin \& Shmakova (2005;2006) describe a similar lensing analysis technique
in which the elements of ${\bf D}$ are referred to as ``Catenoids''
and ``Displacements.''

\begin{figure}[h]
\centerline{\psfig{figure=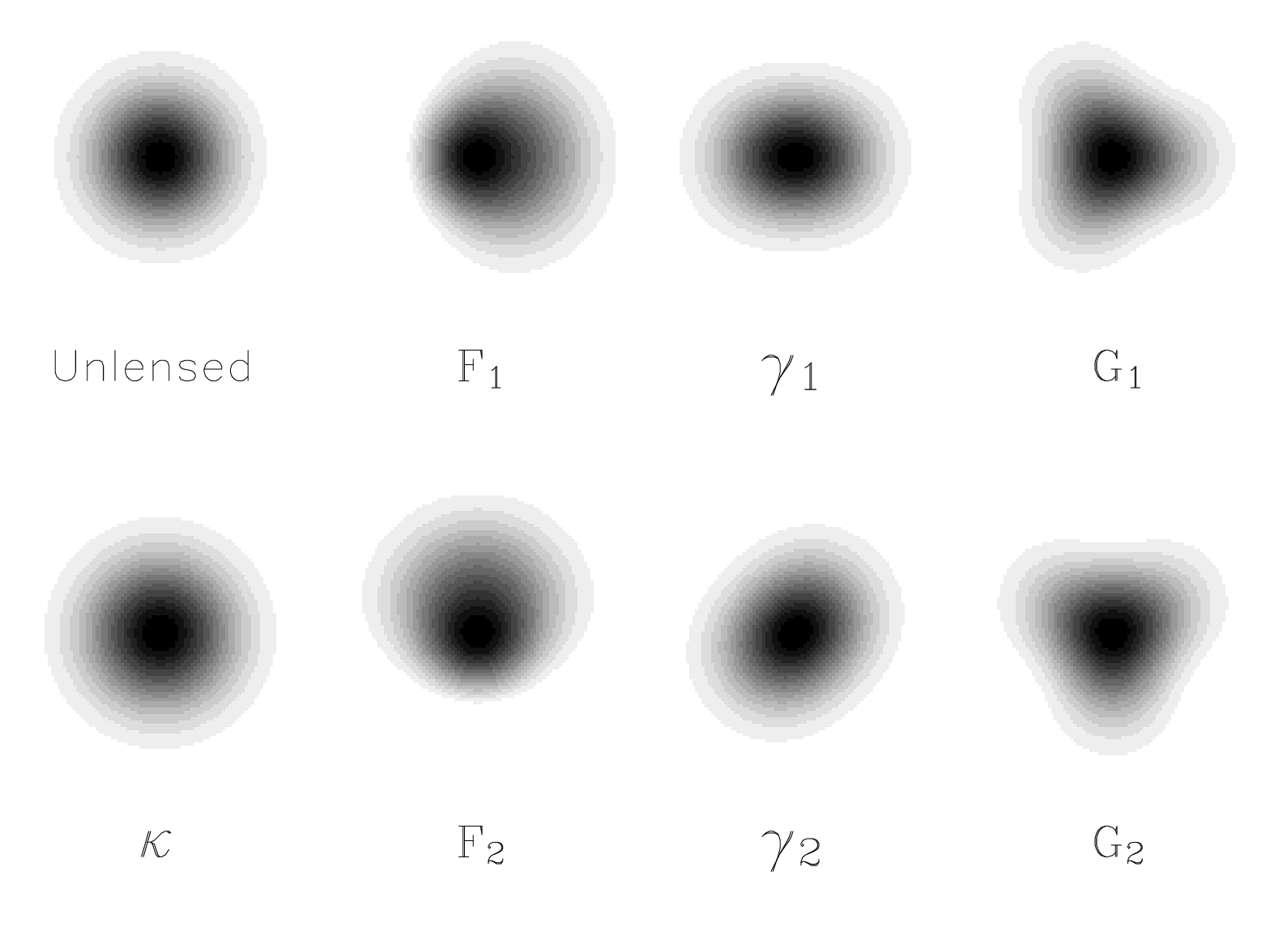,angle=0,height=2in}}
\caption{From Bacon, Goldberg, Rowe \& Taylor, 2006.  The lensed
  images corresponding to pure first or second flexion, shear, or
  magnification.}
\label{fg:shapes}
\end{figure}

BGRT describe an inversion whereby one can estimate the individual
components, and thus measure two ``flexions'':
\begin{equation}
{\cal F}\equiv \partial^\ast \gamma
\end{equation}
\begin{equation}
{\cal G}=\partial \gamma
\end{equation}
where $\partial$ is the complex derivative operator:
\begin{equation}
\partial = \partial_1+i\partial_2\ .
\end{equation}
Figure~\ref{fg:shapes} is reproduced from BGRT and shows the effect of
a first or second flexion on a circular source.

An object with first flexion, ${\cal F}$, appears skewed, while an
object with second flexion, ${\cal G}$, appears arced, especially if
the image has an induced shear as well.  The first flexion has an
$m=1$ rotational symmetry, and thus behaves like a vector.  In
particular, it is a direct tracer of the gradient of the convergence:
\begin{equation}
{\cal F}_1 {\bf i}+{\cal F}_2 {\bf j}=\nabla \kappa
\end{equation}
where ${\cal F}_1$ is the real component and ${\cal F}_2$ is the
imaginary part (as with the second flexion and, as the standard
convention, with shear).

The second flexion has an $m=3$ rotational symmetry, though unlike the
first flexion, it has no simple physical interpretation like that of
the first flexion.  It is, however, roughly proportional to the local
derivative of the magnitude of the shear.  A more complete discussion
of flexion formalism can be found in BGRT.

\section{Shapelets Decomposition}

\label{sec:shapelets}

\subsection{Review of Shapelets}

Measurement of flexion ultimately requires very accurate knowledge of
the distribution of light in an image.  The shapelets (Refregier 2003;
Refregier \& Bacon 2003) method of image reconstruction decomposes an
image into 2D Hermite polynomial bases:
\begin{equation}
f({\bf \theta})=\sum_{n,m} {\cal B}_{nm}({\bf \theta}) f_{nm}\ .
\end{equation}
 This technique has a number of very natural advantages.  In the
absence of a PSF, all shapelet coefficients will have equal noise.
Moreover, the basis set is quite localized (Hermite polynomials have a
Gaussian smoothing filter), and thus is ideal for modeling galaxies.
Furthermore, the generating ``step-up'' and ``step-down'' operators
for the Hermite polynomials are simply combinations of the $x_i$, and
$\partial_i$ operators.

Refregier (2003) shows that if we decompose a source image, $f$, into
shapelet coefficients, the transformation to a lensed image may be
expressed quite simply as:
\begin{equation}
f'=(1+\kappa \hat{K}+\gamma_j \hat{S}_j)f
\end{equation}
where the various lensing operators are:
\begin{eqnarray}
\hat{K}&=&1+\frac{1}{2}\left(\hat{a}_1^{\dagger 2}+\hat{a}_2^{\dagger
  2}-\hat{a}_1^2-\hat{a}_2^2\right)\nonumber \\
\hat{S}_1&=&\frac{1}{2}\left(\hat{a}_1^{\dagger 2}-\hat{a}_2^{\dagger
  2}-\hat{a}_1^2+\hat{a}_2^2\right)\nonumber \\
\hat{S}_2&=&\frac{1}{2}\left(\hat{a}_1^{\dagger}\hat{a}_2^{\dagger}-\hat{a}_1\hat{a}_2\right)\ ,
\end{eqnarray}
$\hat{a}^{\dagger}$ and $\hat{a}$ are the normal step-up
and step-down operators, and the subscript refers to the directional
component of the coefficient (i.e. 1 for the first or x-component, and
2 for the second, or y-component).  Note that in the weak field limit,
these operators indicate that power will be transferred between
coefficients with indices $|\Delta n|+|\Delta m|=2$, which preserves
symmetry as well as keeping the image representation in shapelet space
compact.

In Goldberg \& Bacon (2005), similar (albeit more complicated)
transforms were found relating the derivatives of shear.  We will not
reproduce the full second order operators here, as they are written in
full in the earlier work, but we will point out some key features.
First, some of the elements in the operators have an explicit
dependence on the (unlensed) quadrupole moments of the light
distribution.  This is due to a relatively subtle effect not present
in shear analysis.  Since the flexion signal is asymmetric, the center
of brightness in the image plane will no longer necessarily correspond
to the center of brightness in the source plane, and since the
shapelet decomposition is performed around the center of light, we
need to correct for this.

Most important, though, is the fact that second order lensing terms
yield transfer of power between indices with $|\Delta n|+|\Delta m|=$
1 or 3.  To second order, then, a lensed image can be expressed as:
\begin{equation}
f'=(1+\kappa \hat{K}+\gamma_j \hat{S}_j
+\hat{S}_{ij}^{(2)}\gamma_{i,j})f \ .
\end{equation}

Flexion analysis assumes (as does shear analysis) that the intrinsic
flexion is random, and thus all ``odd'' (defined as n+m) moments are
expected to be zero.  Thus, from a set of shapelet coefficients, a
best estimate for the flexion signal may be found via $\chi^2$
minimization, where:
\begin{equation}
\chi^2\equiv\left[\mu_{n_1m_1}-f_{n_1m_1}+(\gamma_i \hat{S}_i^{(1)}+\gamma_{i,j}S_{ij}^{(2)})
  \overline{f}_{n'_1m'_1}\right]V^{-1}_{n_1m_1 n_2
  m_2}
 \left[(\mu_{n_2m_2}-f_{n_2m_2}+(\gamma_i \hat{S}_i^{(1)}+\gamma_{i,j}S_{ij}^{(2)})
  \overline{f}_{n'_2m'_2}\right],
\label{eq:chi2}
\end{equation}
$V_{n_1m_1 n_2m_2}$ is the covariance matrix of the shapelet
coefficients, and $\mu_{nm}$ is the ``unlensed'' estimate of a shapelet
coefficient.  For odd modes, this is zero.  For even modes, the
relative effect of shear is typically much smaller than the intrinsic
ellipticity of an image, thus it makes sense to set $\mu_{nm}=f_{nm}$.

\subsection{Effective Estimation of the Flexion}

Though the form looks quite complicated, conceptually, computing the
flexion is very straightforward.  A simplified pipeline may be written
as follows:
\begin{enumerate}
\item Generate a catalog of objects and, for each, excise an isolated
  postage stamp.
\item Compute the shapelet coefficients of the postage stamp.
\item Deconvolve the postage stamp with a known PSF kernel.
\item Compute the transformation matrices associated with each of the
  four flexion operators, solve the $\chi^2$ minimization (equation
  \ref{eq:chi2}) for $\gamma_{i,j}$, and estimate the flexion.
\end{enumerate}

We discuss each of these steps in turn below.  The data used for this
analysis was taken using HST and the Advanced Camera for Surveys, and
in the particular context of cluster lensing.  In this context, the
galaxies in which we are interested are potentially blended with much
larger and brighter foreground objects.  We discuss the specific
properties of our data catalog in \S~\ref{sec:measurement}, but many
of the issues involved are quite generic.

\subsubsection{Catalog Generation and Postage Stamp Cutout}

The first step in the process, the generation of a catalog and postage
stamps seems quite straightforward.  For some datasets, such as the
SDSS (York et al. 2000), the data release includes an atlas of pre-cut
postage stamps.  For other applications, such as in relatively shallow
galaxy-galaxy or cosmic shear/flexion studies, fields will be
relatively uncrowded and thus simple application of widely used
packages such as SExtractor (Bertin \& Arnouts, 1996) can be used.

When fields are crowded, however, and contain a wide range of
brightnesses and sizes, the catalog generation becomes more
complicated. It has been noted by Rix et al. (2004) that in general, a
single set of SExtractor parameters is insufficient for detection of
all the objects of interest within an image; setting the source
detection threshold too low will result in excessive blending near
bright objects, whereas a high threshold results in a failure to
detect fainter sources. Rix et al. describe a two-pass strategy for
object detection and deblending involving an initial (``cold'') pass
to identify large, bright objects, followed by a lower-threshold
(``hot'') pass to pick up dimmer objects. Their final catalog consists
of all the objects detected in the cold run, plus any objects detected
in the hot run that do not lie within the isophotal area of any object
detected in the first pass.

This technique works well to prevent spurious deblending by SExtractor
in images in which there is significant substructure. However, when
dealing with crowded fields (such as clusters of galaxies) the largest
problem in catalog generation is excessive blending of sources,
particularly in the central region. To remedy this, we use a modified
version of this hot/cold technique. Our method consists of a primary
SExtractor run to detect only the brightest objects.  In a lensing
field, especially in a lensing cluster, these bright objects will tend
to be the lenses.  Making use of the RMS maps generated during this
SExtractor run, we mask out the bright objects by setting the pixel
values to background noise, and thus simulate an emptier field.

We then run SExtractor on the masked image, using a much lower
detection threshold, to create a catalog of background objects. Since
shape estimation including both flexion and shear have a minimum of 10
degrees of freedom, we require at least 10 connected pixels above the
detection threshold, though in reality, we are unlikely to be able to
get a reliable measurement from an image with fewer than 15 included
pixels.  We then discard all objects for which reliable shape
estimates cannot be found.

%Once sources have been detected in a sample of images, the various
%SExtractor catalogs can be combined into a master catalog. Criteria
%for analysis of any given object in the catalog will be dependent on
%the number and types of images in the data set. For our purposes,
%objects must be detected in at least two different wavelength bands in
%order to be considered for analysis. Additionally, objects should have a
%semi-major axis of at least 1.5 pixels
%to be considered for analysis, and should not be near
%the edge of the image.

For each remaining object, a postage stamp is generated. Ideally, this
should identify any neighboring objects and mask them out (by setting
their pixel values equal to background noise). Our postage stamp code
also identifies objects which are blended by using a
friends-of-friends algorithm to find sets of connected pixels that are
a certain threshold (typically 2-3$\sigma$ for the stacked images
described below) above the background. If there is any overlap between
the object of interest and another object within the field of the
postage stamp, we consider the source to be excessively blended and
exclude it from further analysis.

\subsubsection{Shapelet Decomposition}

Shapelets can be an extremely compact representation of an individual
image.  However, in reality, they are a {\it family} of basis
functions.  There is a characteristic scaling parameter, $\beta$,
which represents the width of the Gaussian kernel in the basis
function Hermite polynomials:
\begin{equation}
{\cal B}_{nm}({\bf \theta})\propto
\exp\left(-\frac{\theta_1^2+\theta_2^2}{2\beta^2}\right) \ .
\end{equation}
In principle, while all values of $\beta$ will yield an orthonormal
basis set, some values produce a dramatically faster reconstruction in
terms of the number of coefficients required to reach convergence.
Moreover, in reality we don't {\it want} to reconstruct all details in
an image.  Structure on the individual pixel scale may simply
represent noise.

From a practical perspective, our goal is to optimize selection of
$\beta$, and the maximum coefficient index, $n_{max}$.  Refregier
(2003) suggests the following parameters:
\begin{eqnarray}
  \beta\simeq\sqrt{\theta_{min}\theta_{max}}\nonumber\\
  n_{max}\simeq\frac{\theta_{max}}{\theta_{min}} -1\ ,
\end{eqnarray}
where $\theta_{min}$ and $\theta_{max}$ represent the minimum and
maximum scales of image structure, respectively.  

R. Massey (private communication) has found that rather than
performing overlap integrals to solve for the shapelet coefficients
(as was done, for example, in the analysis of Goldberg \& Bacon 2005),
the ideal approach is to do a $\chi^2$ minimization of the reconstructed
image with the original postage stamp.  This may seem complicated, and
it is.  Fortunately, a shapelets package is available in IDL at the
shapelets
webpage.\footnote{http://www.astro.caltech.edu/\~{}rjm/shapelets/}

For our sample of co-added, background-subtracted, HST ACS images of
Abell 1689, we find that $\theta_{min} = 0.4$ pixels and $\theta_{max}
= 1.8\sqrt{a^2+b^2}$ give good shapelet reconstructions, where $a$ and
$b$ are the semi-major and semi-minor axes of the galaxy as measured
by SExtractor. However, it is important to note that these parameters
are somewhat dependent on the noise level in the images.

For a sky-limited sample, we have found that the optimal choice of
$\theta_{min}$ scales approximately linearly as the ratio of the flux
to the RMS sky noise.  We have looked at this scaling in a sample of
galaxies detected with ACS (and which we describe in greater detail
below), each of which was imaged in 4 frames.  Prior to stacking of
these frames, we found that $\theta_{min} = 0.75$ produced low
$\chi^2$ and convergence with small values of $n_{max}$.  After
stacking, $\theta_{min}=0.4$ was required. This makes sense, since the
noisier our image, the more prone we might otherwise be to fitting
complex polynomials to what is, essentially, noise.

Roughly, the processing time for a decomposition scales as
$\theta_{max}^4$, as $\theta_{max}$ determines both the postage stamp
size and the maximum order of the shapelet decomposition. Due to the
high resolution of our images, we encountered a number of objects for
which $n_{max}$ was so large that the decomposition time became
prohibitive. We opted to re-grid images with $n_{max} > 50$ into larger
pixels by taking the mean of the pixel values in square bins, the size
of which is determined by $binsize = n_{max}/50$. This number
was rounded up for objects with $50 < n_{max} \le 75$. 

A flexion measurement is then carried out using the $\chi^2$
minimization technique described previously. However, we have found
that truncating the shapelet series prior to the flexion measurement
yields a more accurate and robust measure of the flexion than using
the full series. Excluding the higher order shapelet modes avoids
contamination of the flexion signal by small scale substructure and by
noise (particularly in dimmer objects). We exclude all shapelet modes
with $n > n_{max}/5$ in our flexion measurement. This effectively
increases $\theta_{min}$ to 2 pixels, without affecting the accuracy
of the reconstruction.

\subsubsection{PSF Deconvolution}

One of the complications in measuring properties of lensed images is
that, in practice, they are convolved with a PSF:
\begin{equation}
f({\bf \theta})=\int d^2\theta' P({\bf \theta-\theta'}) f^{(0)}({\bf
  \theta})\ .
\label{eq:PSFdef}
\end{equation}  
In principle, the PSF can be estimated through measurement of stars,
but in deep, small-field, high galactic-latitude observations, stars
may be scarce, and thus PSF estimation may rely partly on numerical
analysis of the instrument (e.g. the Tiny Tim algorithm, Krist,
J. 1993).  

In reality, though, this should rarely be an issue.  Estimations of
the PSF flexion from Tiny Tim yield values of $\sigma_{aF,psf}\simeq
7\times 10^{-4}$.  This represents the maximum induced flexion which
can arise from convolution with the PSF, and is still several orders
of magnitude lower than the scatter in intrinsic flexion of galaxies.

We are not surprised by this since, for example, PSF distortions
arising from variable sizes in chips is likely to scale as the
variation in PSF ellipticity.  In ACS, chip distortions produce
ellipticities of order $1\%$, and vary on scales of 100's of pixels,
producing an induced flexion of $\sim 10^{-4}\ pix.^{-1}$.  From the
ground, the atmospheric distortions are expected, on average, to be
even more isotropic.

There is another reason to suppose that PSF flexion contributions will
be unimportant.  In shear measurements, the PSF ellipticity typically
varies smoothly and somewhat symmetrically around the center of a
field, mimicking (or partially reversing) the overall behavior of the
expected shear field.  Since flexion probes smaller scale effects, the
induced flexion by the PSF will, on average, cancel out.

This is not to say that we cannot deal with PSF flexion inversion.  
Refregier (2003) describes an explicit deconvolution algorithm (see
also Refregier and Bacon 2003, and references therein).  In shapelet
space, equation~(\ref{eq:PSFdef}) can be re-written as:
\begin{equation}
f_{nm} = \sum_{n'm'n''m''} C_{nmn'm'n''m''} P_{n'm'}f^{(0)}_{n''m''}
\end{equation}

Where $C_{nmn'm'n''m''}$ is the 2-dimensional convolution tensor:
\begin{equation}
C_{nmn'm'n''m''}(\gamma,\alpha,\beta)=2\pi(-1)^{n+m}i^{n+m+n'+m'+n''+m''}\int
d^2{\bf x}{\cal B}_{n''m''}({\bf x}/\gamma){\cal B}_{n'm'}({\bf
x}/\alpha){\cal B}_{nm}({\bf x}/\beta)\ ,
\end{equation}
and $\alpha$, $\beta$ and
$\gamma$ are the characteristic scales of $f^{(0)}$, $P$ and $f$,
respectively.  We may then define a PSF convolution matrix as:
\begin{equation}
P_{nmn'm'}\equiv\sum_{n''m''} C_{nmn'm'n''m''}P_{n''m''}\ .
\end{equation}
If only low order terms in the convolution matrix are included, it may
be inverted to perform a deconvolution via:
\begin{equation}
  f^{(0)}_{nm}=\sum_{n'm'} (P^{-1})_{nmn'm'}f_{n'm'}\ .
\end{equation}
This provides a good estimate of the low order coefficients, but high
order information is lost. An alternative inversion scheme involves
fitting the observed galaxy coefficients using a $\chi^2$ minimization
scheme.  Refregier and Bacon (2003) note that the $\chi^2$ scheme may
be more robust numerically, and can take full account of variations in
the noise characteristics across an image (although it is strictly
only valid in the case of Gaussian noise). It is this scheme that is
implemented in the shapelets IDL software.

\subsubsection{Flexion Inversion}

If the shapelet coefficients are statistically independent (as they will
be in the absence of an explicit PSF deconvolution), formal inversion
of the flexion operator is quite straightforward.  Under these
circumstances, we also have the benefit that the measurement error for
each moment is identical (see Refregier 2003 for discussion).  

Noting that, in most galaxies, the coefficients corresponding to the
$n+m=$even moments will be much larger than the odd moments (and,
indeed, upon random rotations, the latter will necessarily average to
zero) we can dramatically simplify equation~(\ref{eq:chi2}).  First,
we define the susceptibility of each odd moment as:
\begin{displaymath}
\Delta f_{n' m',ij}=\hat{S}^{(2)}_{ij}f_{nm}
\end{displaymath}
where $f_{nm}$ represents all of the ``even'' coefficients, and $n'm'$
represents all of the odd coefficients.  Thus, we wish to solve for
the relation:
\begin{equation}
\sum_{n'm'}(f_{n'm'}-\gamma_{i,j}\Delta f_{n' m',ij})^2=min.
\end{equation}
where the first term is taken directly from measurement.  Taking the
derivatives and rearranging, we find:
\begin{equation}
\sum_{n'm'} f_{n'm'}\Delta f_{n'm',ij}=\gamma_{i,j}\sum_{n'm'}(\Delta
f_{n'm',ij}\Delta f_{n'm',i'j'}) 
\end{equation}
which can readily be inverted to solve for $\gamma_{i,j}$.  

In practice, however, there are a number of issues which must be
considered.  First, if the PSF or pixel scale are relatively large
compared to the minimum resolution scale of an image then many of the
high-order moments returned by shapelets decomposition will, in fact,
not have any information.  Thus, the above inversion will yield a
systematic underestimate of the true image flexion.  Above, we
describe a truncation which minimizes this effect.

While the flexion inversion is, at its core, linear algebra, it
involves an enormous number of terms.  We have thus provided an
inversion code for shapelets estimates of flexion along with examples
on the flexion webpage.

\section{HOLICs Analysis}

\label{sec:holics}

\subsection{Higher Order Moments}

Okura et al. (2006) recently related flexion directly to the 3rd
moments of observed images.  This is a significant extension of flexion,
and very much along the lines of Goldberg \& Natarajan's (2002)
original work which talked about ``arciness'' in terms of the measured
octopole moments.  Throughout our discussion, we will use the
notation:
\begin{equation}
Q_{ij}=\frac{1}{F}\int (\theta_i-\overline{\theta_i}) (\theta_j-\overline{\theta_j}) f({\bf
  \theta}) d^2{\bf \theta}
\end{equation}
to refer, in this case, to the unweighted quadrupole moments, with all
higher moments being defined by exact analogy.  In this context, $F$
refers to the unweighted integrated flux.

They define the complex terms:
\begin{equation}
\zeta\equiv\frac{(Q_{111}+Q_{122})+i(Q_{112}+Q_{222})}{\xi}
\end{equation}
and
\begin{equation}
\delta\equiv\frac{(Q_{111}-3Q_{122})+i(3Q_{112}-Q_{222})}{\xi}
\end{equation}
where
\begin{equation}
\xi\equiv Q_{1111}+2Q_{1122}+Q_{2222}\ .
\end{equation}
These terms are collectively referred to as HOLICs.

If a galaxy is otherwise perfectly circular (i.e. no ellipticity), and
in the absence of noise, then the HOLICs may be directly related to
estimators of the flexion (subject to an unknown bias of
$1-\kappa$).  Namely:
\begin{equation}
{\cal F}\simeq \frac{4 \zeta\xi}{9\xi-6(Q_{11}^2+Q_{22}^2)}
\label{eq:skewness}
\end{equation}
\begin{equation}
{\cal G}\simeq \frac{4 \delta}{3}
\label{eq:arciness}
\end{equation}
where the latter term in the denominator of ${\cal F}$ does not appear
in the Okura et al. analysis.  Bacon and Goldberg (2005) show that a
flexion induces a shift in the centroid proportional to the quadrupole
moments.  In order to correctly invert the HOLICs, this term needs to
be incorporated explicitly.  The simplicity of the extra term results
from an approximation of near circularity.

The beauty of this approach is that it gives us a very intuitive feel
for what flexion means in an observational way.  We thus introduce the
term ``skewness'' to the intrinsic properties of a galaxy as measured
from equation~(\ref{eq:skewness}) whether or not the galaxy is otherwise
circular, and whether or not it is lensed.  The skewness may be
thought of as the intrinsic property, much as the ``ellipticity'' is
the intrinsic property related to the ``shear.''  Likewise, the
intrinsic property associated with equation~(\ref{eq:arciness}) will be
referred to as the ``arciness.''

In reality, however, equations~(\ref{eq:skewness}) and
(\ref{eq:arciness}) are not sufficient to perform a flexion estimate
even if a galaxy has an ellipticity of only a few percent.  Okura et
al provide a general relationship between estimators for flexion and
HOLICs, though the relation is best expressed in matrix form:
\begin{equation}
{\cal M} \left(
\begin{array}{c}
{\cal F}_1 \\
{\cal F}_2 \\ 
{\cal G}_1 \\
{\cal G}_2
\end{array}
\right)
=
\left(
\begin{array}{c}
\zeta_1 \\
\zeta_2 \\ 
\delta_1 \\
\delta_2
\end{array}
\right)
\label{eq:Fsolve}
\end{equation}
where ${\cal M}$ is a $4\times 4$ matrix consisting of elements
proportional to sums of $Q_{ijkl}$ and $Q_{ij}Q_{kl}$, the former of
which can be found by explicitly expanding the expressions in Okura et
al., and the latter of which is again derived from the shift in the
centroid.  For the convenience of the reader, we write out the
explicit form of ${\cal M}$ in Appendix A.

It may be seen by examining the elements of ${\cal M}$ why this
inversion must be done explicitly for even mildly elliptical sources.
For fully circular sources, it may be seen by inspection that ${\cal
M}$ is diagonal.  However, when a source has an ellipticity even as
small as $10\%$, it can be shown that $|M_{11}|\simeq |M_{12}|$, and
thus equations~(\ref{eq:skewness}) and (\ref{eq:arciness}) are no
longer even approximately correct.

\subsection{Gaussian Weighting with HOLICs}

The application of the HOLICs technique would be trivial if there were
no measurement noise.  In the presence of noise, and especially, when
the sky dominates, measurement of unweighted moments is inherently
quite noisy.  In a case where we are measuring the 3rd and 4th
moments, it is even more so.

Kaiser, Squires \& Broadhurst (1995; see also a nice review by
Bartelmann \& Schneider, 2001) developed perhaps the most
comprehensive approach to dealing with the second moments (the
ellipticity) with noisy observing, and with a (potentially
anisotropic) PSF.

Our approach is similar.  We have only worked with a Gaussian window
thus far, but the approach is generalizable for any circularly
symmetric weighting.  We thus define a window function:
\begin{equation}
W({\bf
  \theta})=\frac{1}{2\pi\sigma_{W}}\exp\left(-\frac{\theta_1^2+\theta_2^2}{2\sigma_W^2}\right)
\end{equation}
where the origin is taken to be the center of light, and the integral
is normalized to unity.  Further, we
define the weighted moments as, for example:
\begin{equation}
\hat{Q}_{11}=\frac{1}{\hat{F}}\int (\theta_1-\overline{\theta_1})^2 f({\bf
  \theta}) W({\bf \theta}) d^2{\bf \theta}\ .
\end{equation}
We can thus redefine all HOLICs and moments similarly.  We have found
through experimentation (see below) that for a sky noise limited
source, a reasonable value of $\sigma_W$ is 1.5 times the half-light
radius.

If we were to simply replace all elements in ${\cal M}$, $\zeta$, and
$\delta$ from equation~(\ref{eq:Fsolve}) with their weighted
counterparts we would not get an unbiased estimate of the flexion.
There are two corrections.  One has to do with the fact that centroid
shift will differ from the unweighted case to the weighted case.
Consider an extreme scenario in which the window width is arbitrarily
small and in which the unlensed image was circularly symmetric with a
peak at the center.  In that case, the centroid will essentially
remain at the center (peak brightness) even if the unweighted moments
shift.

Thus, compared to the unweighted moments, the centroid will shift:
\begin{equation}
\Delta \overline\theta_l=\frac{\sum_{ijk}D_{ijk}\hat{Q}_{ijkl}}{\sigma_W^2}
\end{equation}
where we have used the explicit fact that for a Gaussian:
\begin{equation}
\frac{dW({\bf \theta})}{d\theta_i}=-\frac{\theta_i}{\sigma_W^2}W({\bf
    \theta})\ ,
\end{equation}

The other correction has to do with the fact that though lensing
preserves surface brightness, it does not preserve total flux.  This
is normally related by the Jacobian of the coordinate transformation.
However, when considering a window function, we need consider that
transformation explicitly:
\begin{equation}
W(\beta)d^2\beta=\left|\frac{\partial \beta}{\partial
  \theta}\right|W(\beta)d^2\theta
\end{equation}
as used by Okura et al. (2006), and where we have simply multiplied
both sides by the factor $W(\beta)$.  In this context, ${\bf \beta}$
refers to the image coordinate in the source plane.  Ignoring the terms
proportional to shear (which cannot be directly addressed by this
method at any rate), we have the approximate relation:
\begin{equation}
W(\beta)\simeq W(\theta)+\frac{1}{2}D_{ijk}\theta_i\theta_j
\frac{dW}{d\theta_k}
\end{equation}
or, as we have already asserted:
\begin{equation}
W(\beta)\simeq
W(\theta)-\frac{1}{2}D_{ijk}\frac{\theta_i\theta_j\theta_k}{\sigma_W^2}
W(\theta)\ .
\end{equation}
Note that the latter term contains an odd number of position elements,
and thus, coupling to the generating equations for $\zeta$ and
$\delta$ produces contributions of $6^{th}$ moments in ${\cal M}$:
\begin{equation}
\Delta
\hat{Q}_{ijk}=-\frac{1}{2}D_{lmn}\frac{\hat{Q}_{ijklmn}}{\sigma_W^2}\ ,
\end{equation}
which, in turn, must be corrected for.

We may thus say that:
\begin{equation}
\hat{\cal M}={\cal M}(\hat{Q}_{ij},...)+\Delta{\cal M}\ ,
\end{equation}
where the latter expression can also be found in Appendix A.  

\subsection{PSF Correction in HOLICs}

\label{sec:psfcorrection}

As with our discussion of shapelets, above, we must also consider PSF
deconvolution in our HOLICs pipeline.  We define the PSF function in
equation~(\ref{eq:PSFdef}), and all unweighted moments of the PSF are
denoted by $P_{ij}$, etc. In principle, because of the higher
signal-noise of the PSF, the unweighted moments are easier to estimate
than the moments of the detected image.  While we argued, above, that
the induced flexion from a PSF is likely to be small, it is still the
case, as with shear, that the PSF will reduce the measured flexion.
Let's first consider the case in which we were able to measure the
unweighted moments of both the PSF and the observed image.  It is
straightforward to show that:
\begin{equation}
Q_{ij}=Q_{ij}^{(0)}+P_{ij}\ .
\label{eq:quadrel}
\end{equation}

Thus may be computed via the relation:
\begin{equation}
Q_{ij}=\frac{1}{F}\int \theta_i \theta_j f^{(0)}(\theta')P(\theta-\theta')
d^2\theta d^2\theta'
\end{equation}
Making the substitution, 
\begin{equation}
\theta''=\theta-\theta'
\end{equation}
yields 
\begin{equation}
Q_{ij}=\frac{1}{F}\int (\theta_i'\theta_j'+\theta_i''\theta_j''+
\theta_i'\theta_j''+ \theta_i''\theta_j')f^{(0)}(\theta')P(\theta'')
d^2\theta'd^2\theta''
\end{equation}
It is straightforward to show that this yields
equation~(\ref{eq:quadrel}).

Similarly, it may be shown that:
\begin{equation}
Q_{ijk}=Q_{ijk}^{(0)}+P_{ijk}\ .
\end{equation}
However,
\begin{equation}
Q_{1111}=Q_{1111}^{(0)}+P_{1111}+6Q_{11}^{(0)}P_{11}\ ,
\end{equation}
with a similar expression for $Q_{2222}$, and 
\begin{equation}
Q_{1122}=Q_{1122}^{(0)}+P_{1122}+Q_{11}^{(0)}P_{22}+Q_{22}^{(0)}P_{11}
\end{equation}
provided we assume the PSF is nearly circular.

If we further look only at nearly circular sources, then we may
estimate the flexion using the forms in equations~(\ref{eq:skewness})
and (\ref{eq:arciness}).  Again, assuming unweighted moments, and zero
PSF and intrinsic flexion we find:
\begin{equation}
\tilde{{\cal F}}_i={\cal F}_i
\frac{9\xi^{(0)}-6(Q^{(0)2}_{11}+Q^{(0)2}_{22})}{9\xi-6(Q_{11}^2+Q_{22}^2)}
\label{eq:fcorrect}
\end{equation}
Where ${\cal F}_i$ is an unbiased estimate of the flexion, and
$\tilde{{\cal F}}_i$ is the estimated flexion if one does not include
the correction for the PSF.  However, the normalization constant may
be estimated directly from combinations of the PSF 2nd and 4th
moments, and the unweighted moments of the image.  Since this term
represents something like the overall radial profile of the source,
the unweighted moments can be estimated even under noisy conditions.

Similarly, the second flexion may be estimated as:
\begin{equation}
\tilde{{\cal G}}_i={\cal G}_i
\frac{\xi^{(0)}}{\xi}
\end{equation}

Though we have derived these relations for a nearly circular source,
we have found they provide a good correction even when the PSF and
intrinsic image size are comparable, and when ellipticities for the
source image are $\varepsilon\simeq 0.2$.

\section{Simulated Lensing}

\label{sec:simulate}

Which approach is better, shapelets or HOLICs?  From a signal
perspective, the shapelets technique is better.  It is designed to
provide optimal weighting and return optimal signal-noise. Moreover,
as described above, inversion of the PSF is a straightforward and
well-designed process.  In the absence of noise, the two techniques
produce very similar results.

On the other hand, the HOLICs technique has several practical
advantages, especially for large surveys.  For one, the HOLICs code is
typically much faster than shapelets.  For an N pixel image, the
HOLICs technique is an ${\cal O}(N)$ calculation, whereas the
shapelets is ${\cal O}(N^2)$. Additionally, some values of $\beta$
produce very bad reconstructions, and hence, minimization of $\chi^2$
can be time-consuming and may not converge to a minimum.

As a simple test, we created images with brightness profiles of:
\begin{equation}
I(r)\propto \exp[-(r/r_0)^n]
\end{equation}
and though we found similar results for a reasonable range of
exponents, the results presented below are for $n=1.5$.  We have used
a constant source ellipticity, typical of those observed in the field,
$\varepsilon=0.2$, and had measurement errors which were dominated by
sky brightness.  In each case, we had no intrinsic arciness or
skewness (that is, the flexion of the unlensed objects were zero),
since our aim was to measure the response of each of the estimators to
lensing.

We then artificially lensed each of our simulated images, added sky
noise, and measured the flexion using both the HOLICs and shapelets
techniques.  The noise is fixed throughout this discussion, as is the
strength of the flexion signal.  It is clear, however, that all
relevant signal-noise values will scale linearly with the strength of
the lensing signal and inversely with sky noise.  

\subsection{Optimzing the HOLICs Scale Factor}

\begin{figure}[h]
\centerline{\psfig{figure=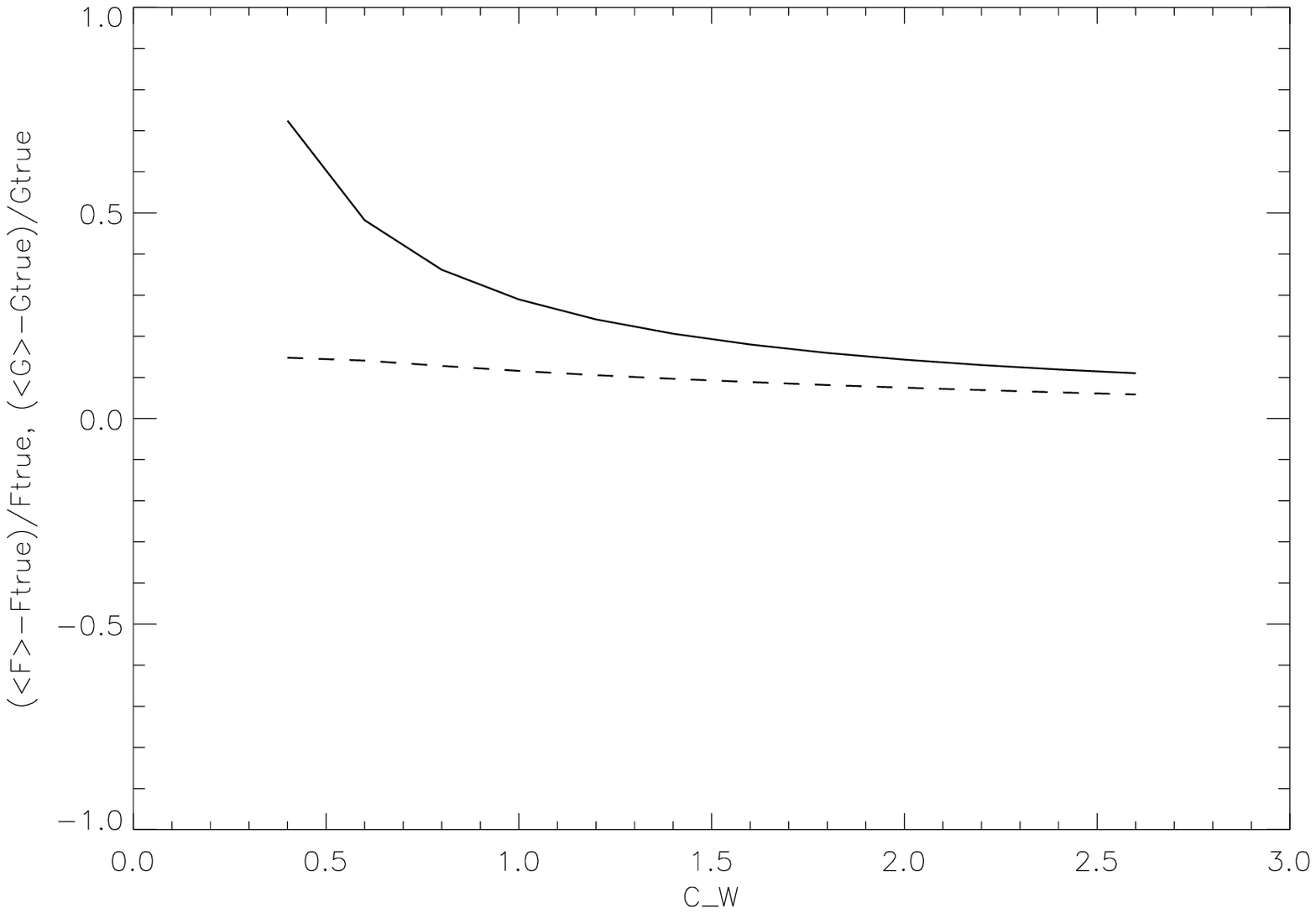,angle=0,height=2.5in}
  \psfig{figure=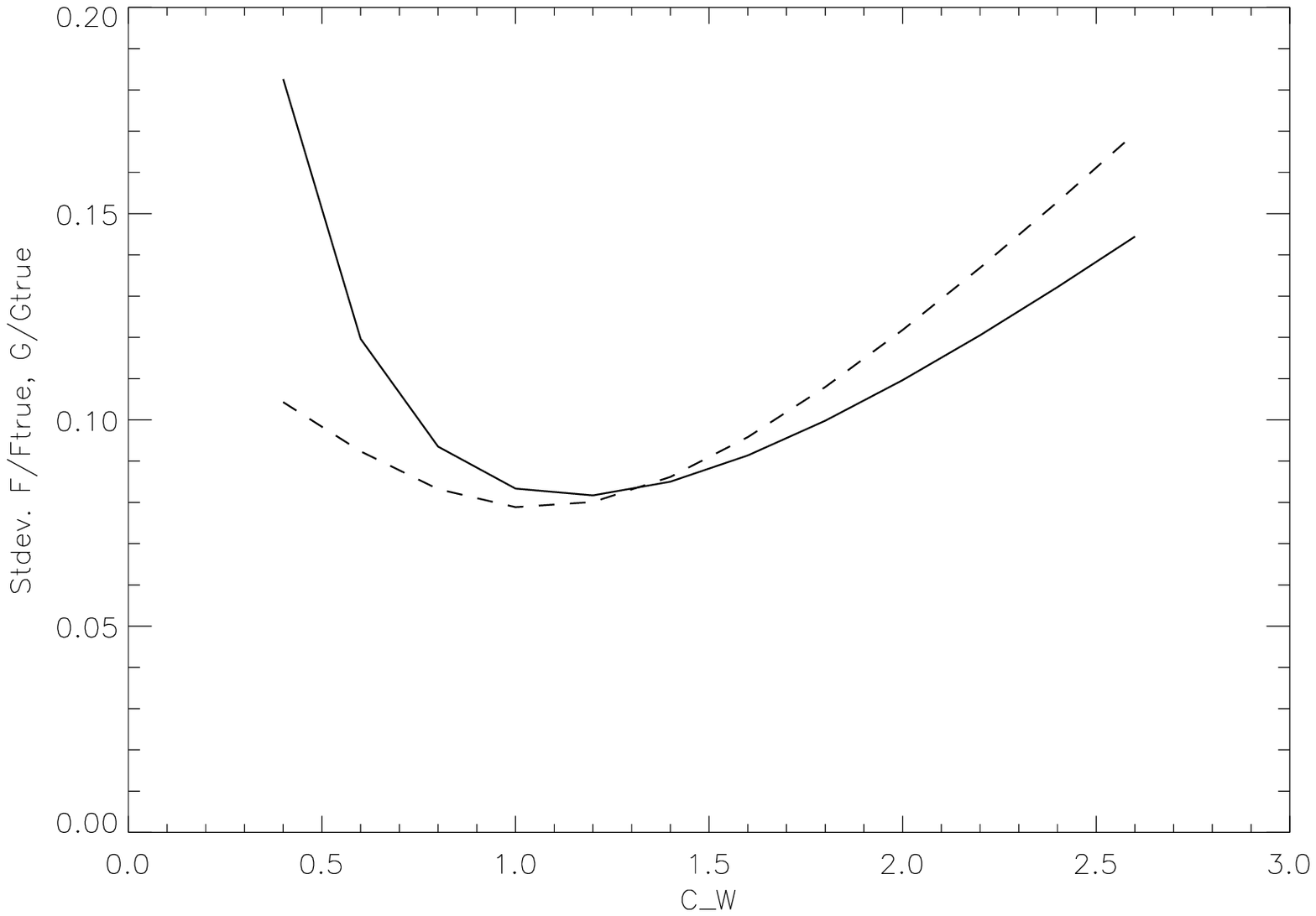,angle=0,height=2.5in}}
\caption{The fractional error in the mean (left panel) and standard
  deviations (right panel) of the measured, simulated 1st flexion
  (solid), and 2nd flexion (dashed), as discussed in the text.}
\label{fg:findC}
\end{figure}

Our first questions is, what is the optimal value of $C_W$, such that:
\begin{equation}
\sigma_W=C_W\times r_{half-light} \ ?
\end{equation}
Ideally, we would like an unbiased estimator of the flexion which also
has very little scatter.  It is clear that the larger the value of
$C_W$, the larger the scatter will be (in general), since we will be
measuring more and more of the noisy sky.  However, the smaller the
$C_W$, the less accurate will be our measure of the real shape of the
galaxy.  Figure~\ref{fg:findC} bears this out.  There is an optimal
value of $C_W$ around 1.5, which reflects a balance between minimizing
measurement errors as well as any measurement bias inherent in the
technique.

With shapelets, we find a systematic underestimate of 11\% in the
first flexion, and an overestimate of 12\% in the second flexion.  We
find a scatter of about 12\% in both.  This is very similar in
magnitude to the results found by an ``optimal'' HOLICs analysis.

\subsection{Correlation of HOLICs and Shapelets Measurement Error}

\begin{figure}[h]
\centerline{\psfig{figure=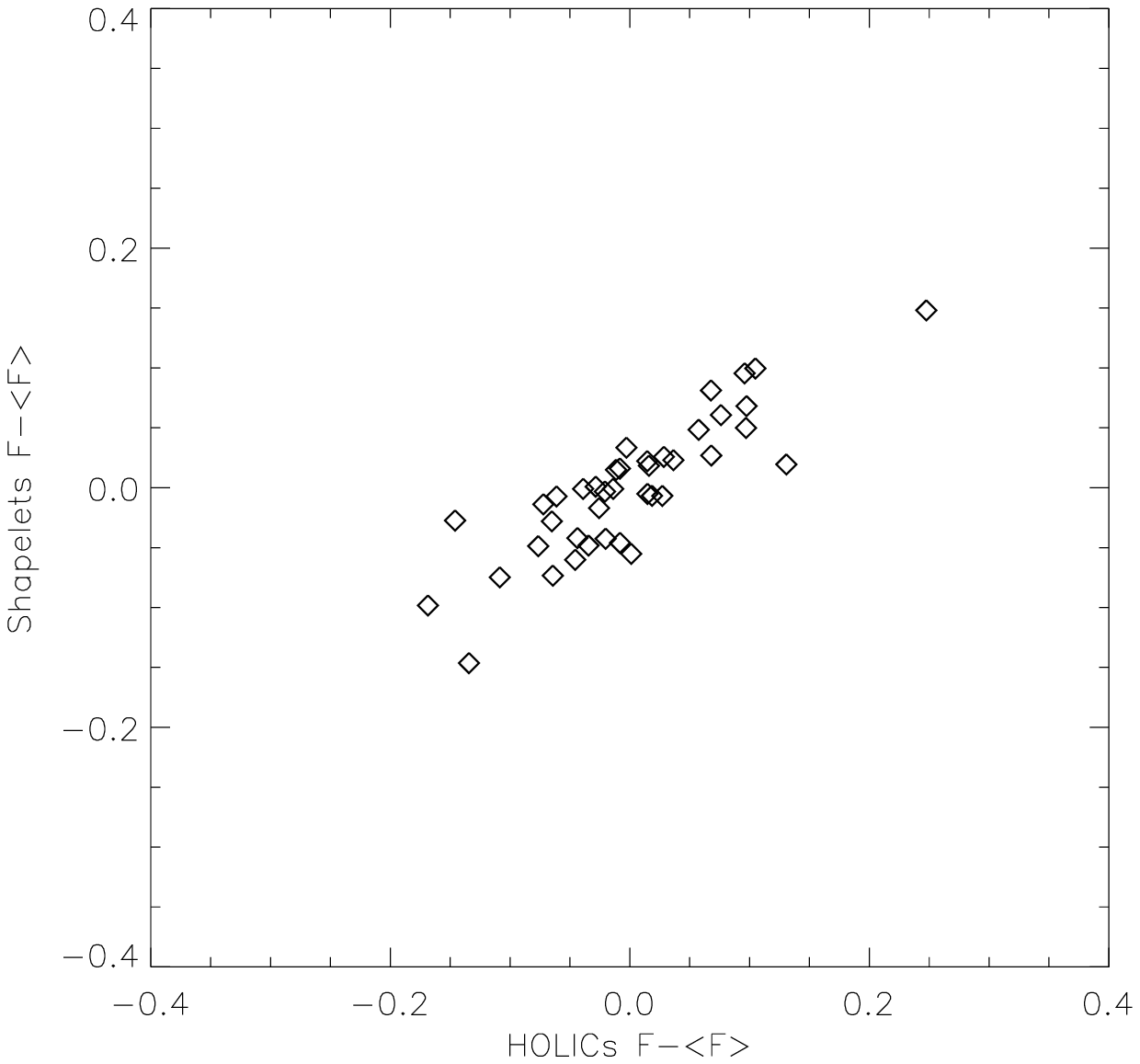,angle=0,height=2.5in} \psfig{figure=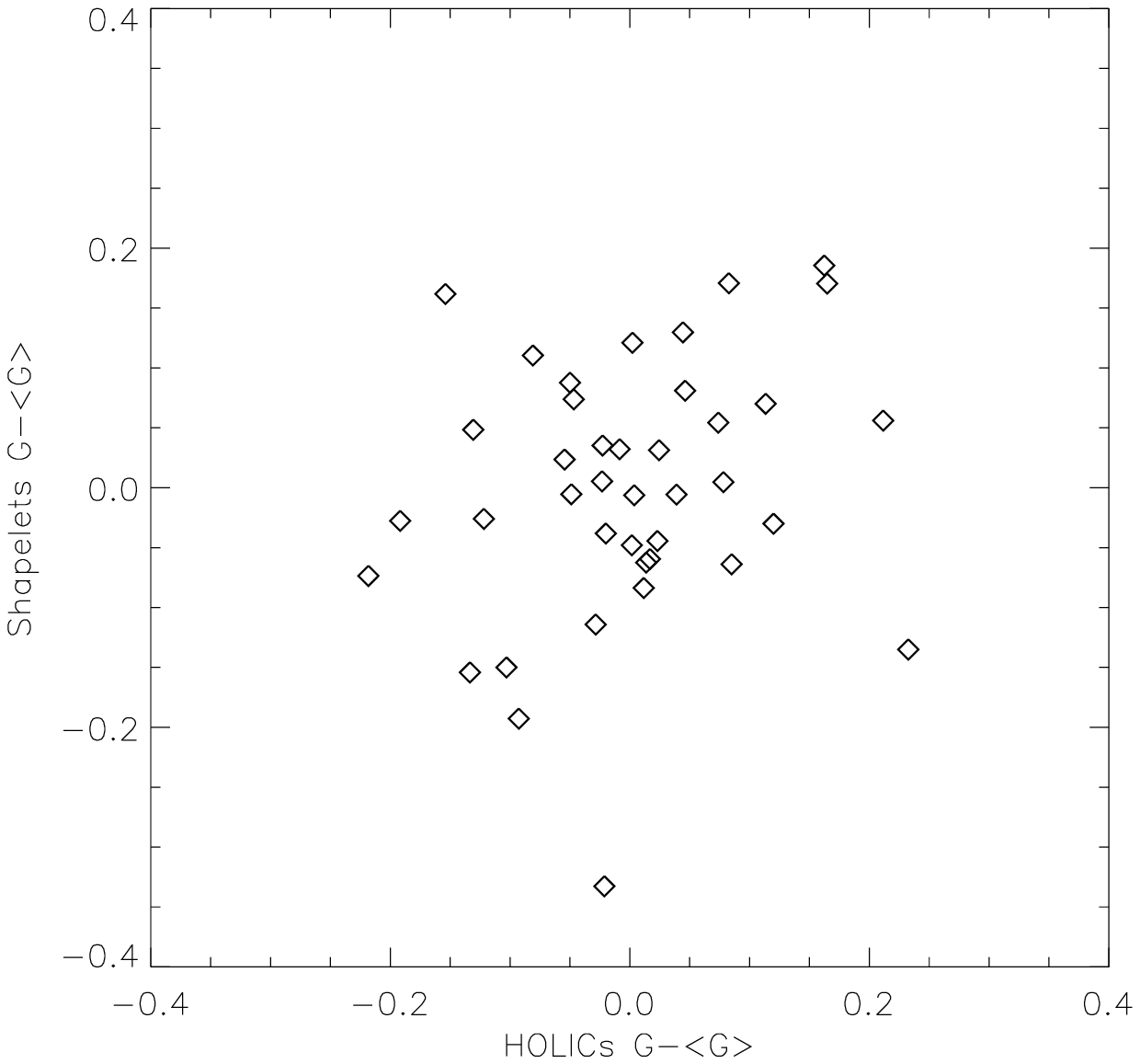,angle=0,height=2.5in}}
\caption{Scatter plots showing the fractional random errors in first
  flexion (left panel), and second flexion (right panel) from HOLICs
  (x-axis), and shapelets (y-axis).  The correlation is quite high for
  1st flexion, with a Pearson correlation coefficient of
  $\rho_F=0.86$, and $\rho_G=0.23$.
}
\label{fg:corrsim}
\end{figure}

Since both HOLICs and shapelets give similar measurement errors at
fixed sky noise, it is worth considering whether we expect measurement
errors between the two techniques to be correlated.  Even in these
idealized circumstances, uncorrelated errors would mean that there is
significant information in the images which is not being used.  In
Fig.~\ref{fg:corrsim}, we show the correlation in uncertainty between
our $C_W=1.5$ HOLICs estimates, and our shapelets estimates.

For the first flexion, in particular, the correlation is quite high,
with a Pearson correlation coefficient of $0.86$.  The correlation in
measurements of the second flexion is much lower, with $\rho_G=0.23$.
Why don't they have perfect correlation?  The two techniques weight
various components of the signal (and thus, the noise) differently,
and therefore have a slightly different response to the noise.  

This general trend is borne out with observed objects as well, in
which we will see much higher correlation between measurements of the
first flexion than the second flexion between the two techniques.

\subsection{PSF Deconvolution}

\begin{figure}[h]
\centerline{\psfig{figure=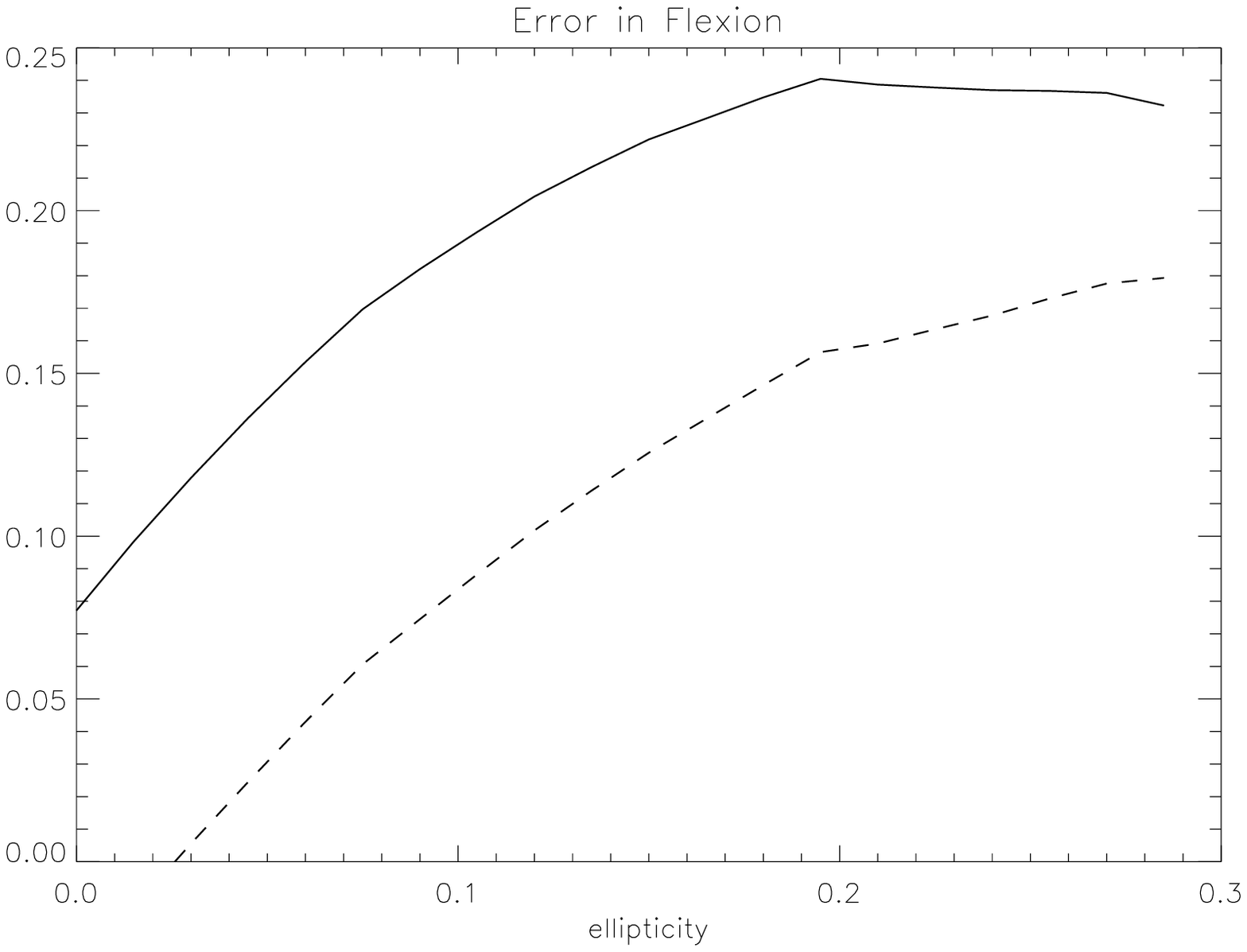,angle=0,height=2in}}
\caption{The fractional error in the recovered flexion estimate using
  a Gaussian PSF with scale of order the size of the intrinsic image.
  Note that errors increase with increasing ellipticity, but produce
  reasonable errors ($\sim 25\%$) in both the 1st flexion (solid
  line), and 2nd flexion (dashed).  No measurement noise is included
  in these calculations, and thus, this result represents the
  systematic error in measurement.}
\label{fg:psf}
\end{figure}

Finally, we can simulate PSF deconvolution.  Using a Gaussian PSF with
a characteristic size somewhat larger than the intrinsic image (the
correction factor described in equation~\ref{eq:fcorrect} is 2.7), we
distorted and then recovered the flexion estimates from images of
increasing intrinsic ellipticity.  This analysis is done in the
absense of sky noise, and thus any errors in shape recovery represent
a systematic effect.  We show the fractional errors in measurement of
the first and second flexion in Figure~\ref{fg:psf}.  Since it is
possible to estimate the systematic error for a combination of
measured shear and PSF shape, it is advisable to those wishing to make
high-precision flexion measurements to take this empirical correction
into account.

We find that, despite the fact that the PSF correction is based on an
assumption of circularity, it continues to produce a good result even
if the image has an intrinsic ellipticity as high as 0.3.

\section{Measurement of Flexion on HST Images}

\label{sec:measurement}

\subsection{Sample Selection and Pipeline}

We also compare the two approaches to flexion inversion on real
objects.  Our data consists of 4 HST ACS cosmic-ray rejected (CRJ)
images of Abell 1689 using the F625W WFC filter (hereafter
``R-band'').  Each image was taken by H. Ford during HST Cycle 11, and
has an exposure time between 2300-2400 seconds.  The observations are
described in detail in Broadhurst et al. (2005).

Using the SWarp software package
\footnote{http://terapix.iap.fr}, these four images were co-added to
create a single ``full'' R-band image. We also generated 2 independent
``split'' images for comparison purposes by combining only two of the
original images. The images are background-subtracted, aligned and
re-sampled, then projected into subsections of the output frame using
a gnomonic (or tangential) projection, and combined using median pixel
values.

Each image undergoes a primary SExtractor run designed to detect only
the foreground objects (cluster members and known stars). This
detection is carried out using the cross-correlation utility in
SExtractor, which allows us to specify the locations of the foreground
objects. Our foreground object catalog was generated using a
combination of spectroscopically confirmed cluster members (Duc et
al. 2002), and identification by eye of foreground objects that were
later confirmed as such by use of the NASA/IPAC Extragalactic Database
(NED), as well as clearly identifiable stars in the field. These
objects are then masked out as described previously, and a second
SExtractor run carried out.

A catalog of objects is then generated, using only those objects that
were detected in both of the split R-band images. We measure the
flexion in our catalog objects using both the truncated shapelets
method (described above) and the HOLICs approach, and then compare the
measurements by computing Pearson correlation coefficients between the
different estimates in the full image. We also compute correlation
coefficients between measurements taken using the same technique in
the two split images. This gives us an estimate of the robustness of
the measurement technique.

When computing the correlation coefficients, we include only objects
with $a > 3$ pixels, and consider only the brightest half of our
catalog objects. In order to exclude extreme or erroneous
measurements, we require $(a|\cal{F}|) <$ 0.2 and $(a|\cal{G}|)<$ 0.5.

\subsection{Results}

\begin{figure}[h]
\centerline{\psfig{figure=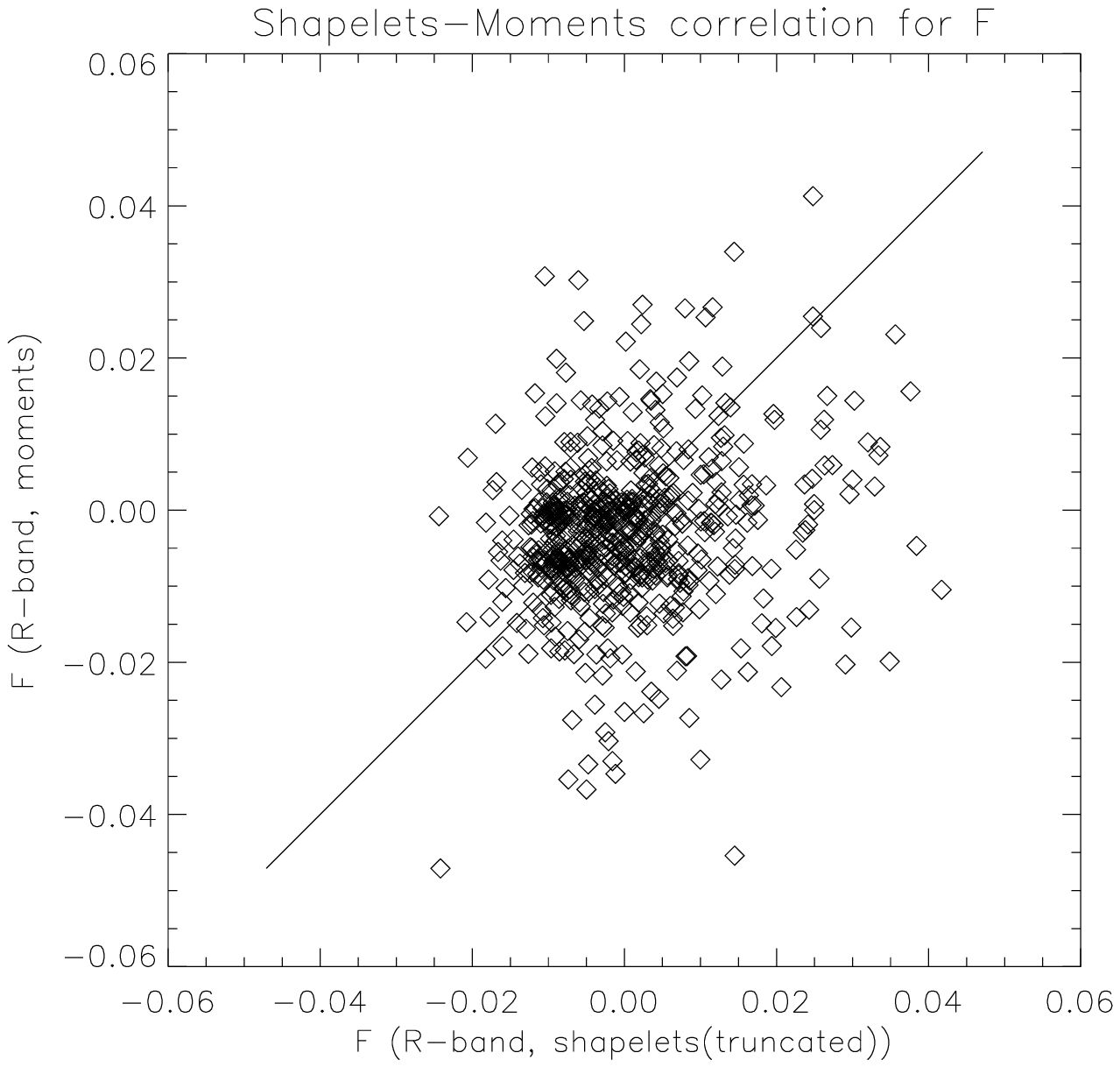,angle=0,height=2.5in} \psfig{figure=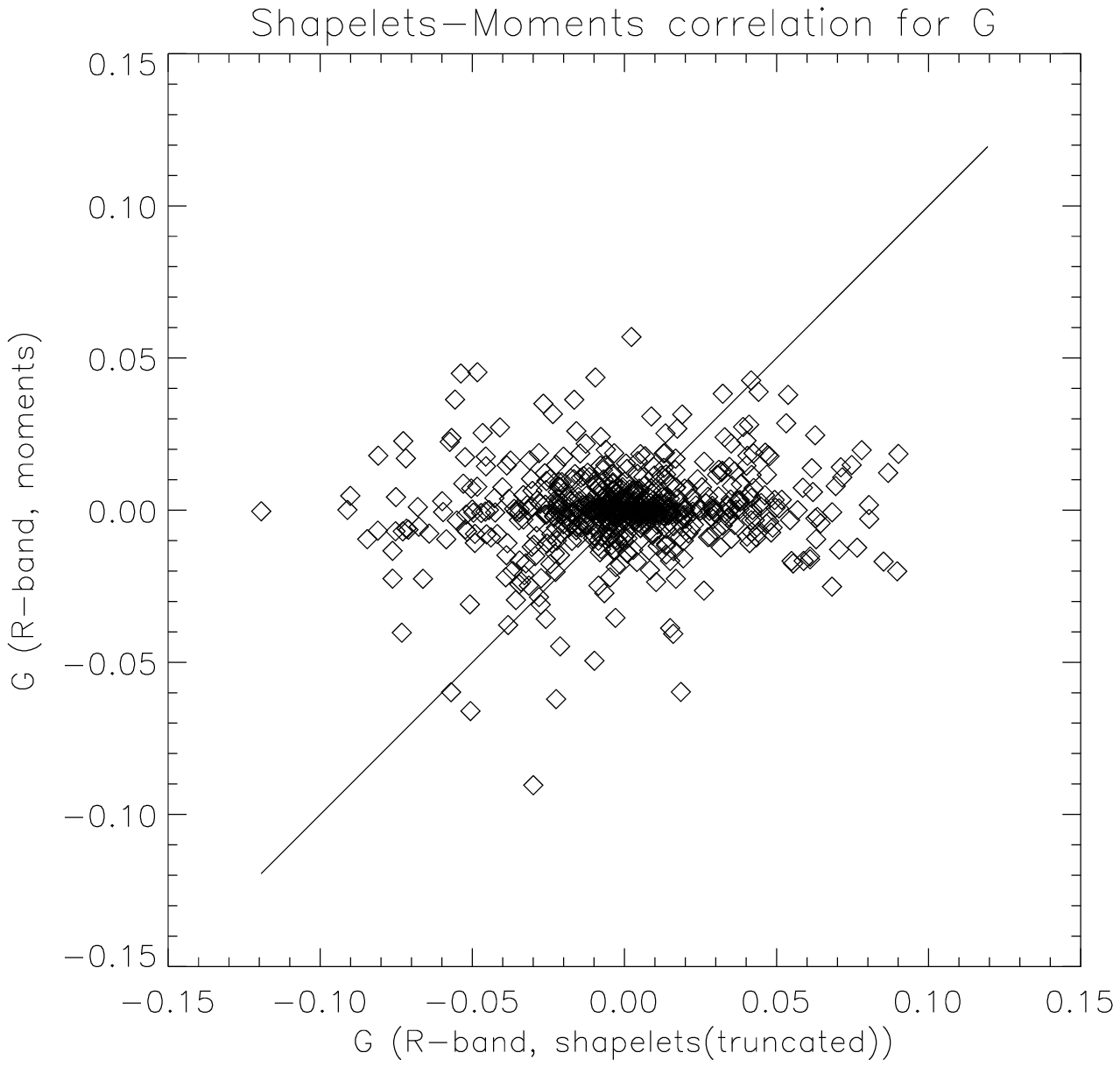,angle=0,height=2.5in}}
\caption{A scatter plot of the values of ${\cal F}$ and ${\cal G}$ as measured using a
  truncated shapelets technique and a HOLICs (or moments) method in the
  full stacked image. }
\label{fullcorr}
\end{figure}

Figure~\ref{fullcorr} shows a comparison of the HOLICs and shapelets
estimates of flexion in the full image. Both $\cal{F}$ and $\cal{G}$
have a positive correlation, with a Pearson correlation coefficient of
0.17 for $\cal{F}$ and 0.12 for $\cal{G}$. Additionally, both methods
yield similar standard deviations for both first and second flexion.

This is what we expected from our simulated results above.  Clearly,
if flexion represents any real signal, the two techniques should be
correlated, and, as we showed in our simulated results, the
correlation in first flexion is higher than in second flexion.  But
the correlation in our measured results is lower than in the simulated
ones.  Why?  In part, this is due to a relatively noisy field.  We've
found that selections on brighter magnitudes and larger objects
improves the correlation somewhat.  In part, however, this is due to
what we mean by ``flexion.''  Recall that the shapelets and HOLICs
analysis of flexion involve weighting different modes in different
ways.  Real, unlensed, galaxies will have odd modes which are not
necessarily correlated in a simple or obvious way.  Lensing, of
course, produces a significant correlation, and thus, a population of
significantly lensed objects (for which the majority of the flexion is
due to lensing) would be expected to have a more correlated flexion.
This is similar to the case with weak shear analysis in that the S/N
from a typical object is usually less than 1.

We can test this hypothesis directly by comparing the measurements in
the split images and estimating the flexion in both using the same
technique.  Any discrepancies between the two ought to be the result
of photon noise rather than intrinsic complexity in the structure of
the 3rd moments.

\begin{figure}[h]
\centerline{\psfig{figure=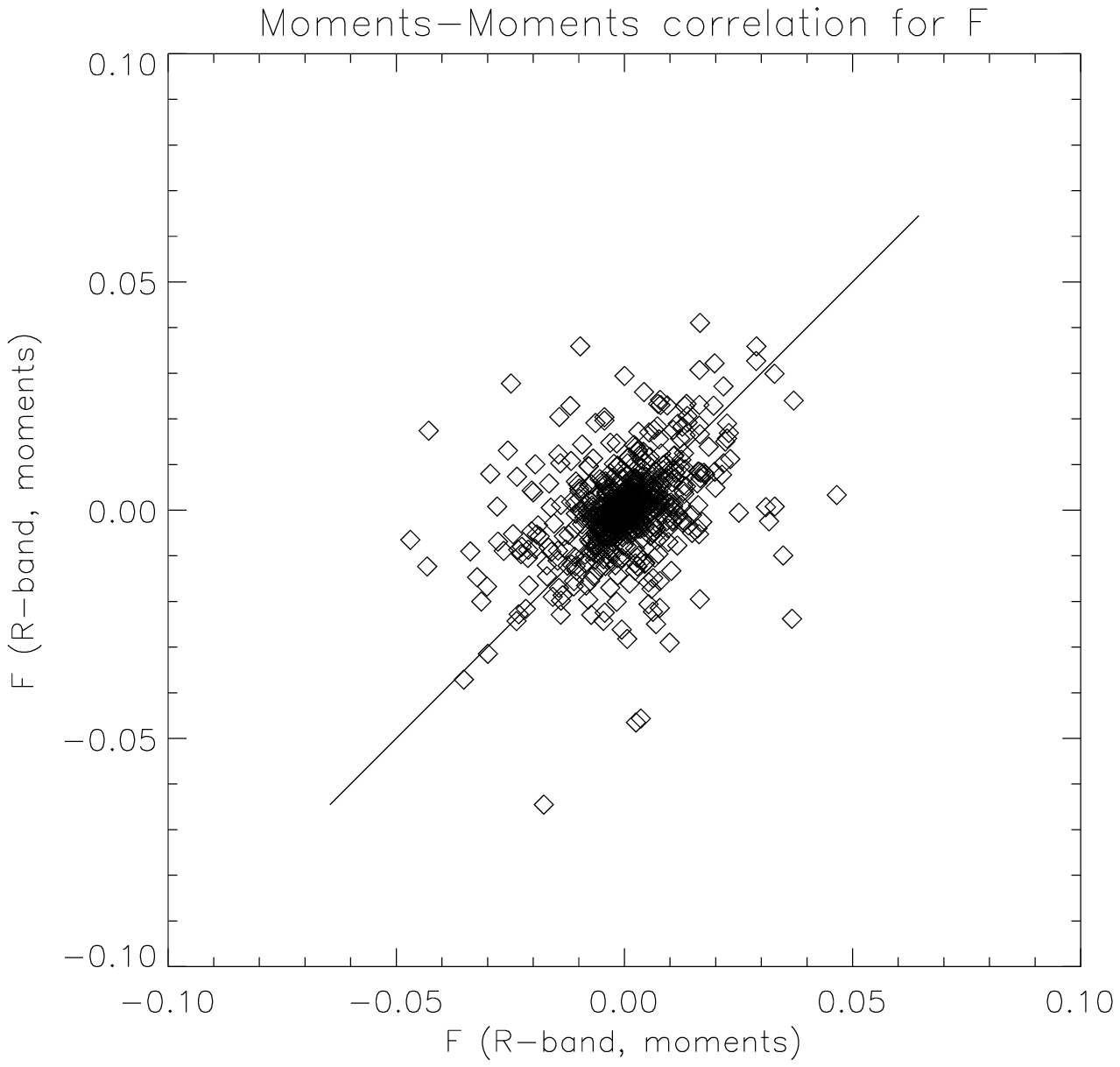,angle=0,height=2.5in}
  \psfig{figure=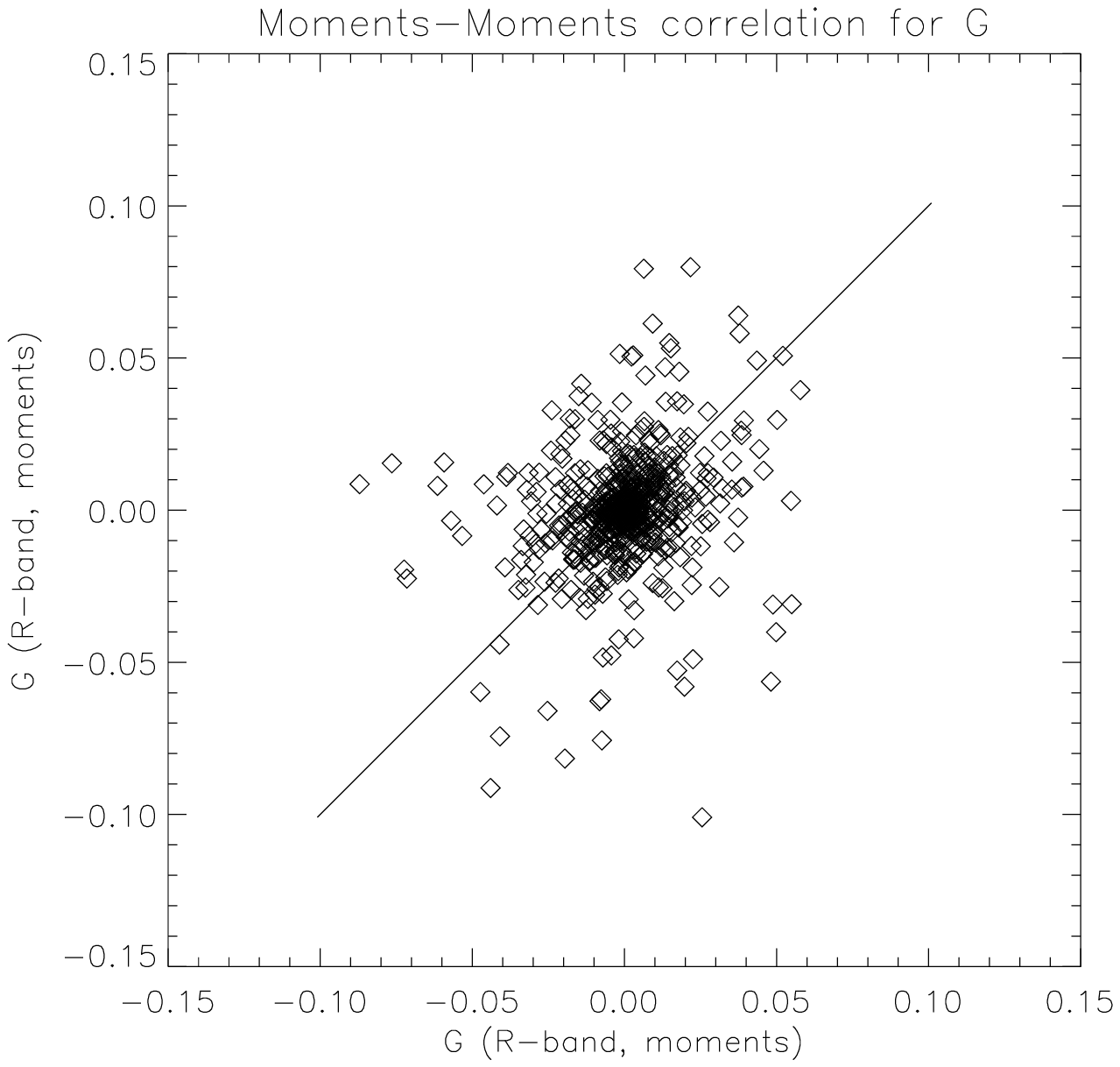,angle=0,height=2.5in}}
\caption{A comparison of HOLICs measurements of ${\cal F}$ and ${\cal G}$ in the split
  images.}
\label{split_moments}
\end{figure}

\begin{figure}[h]
\centerline{\psfig{figure=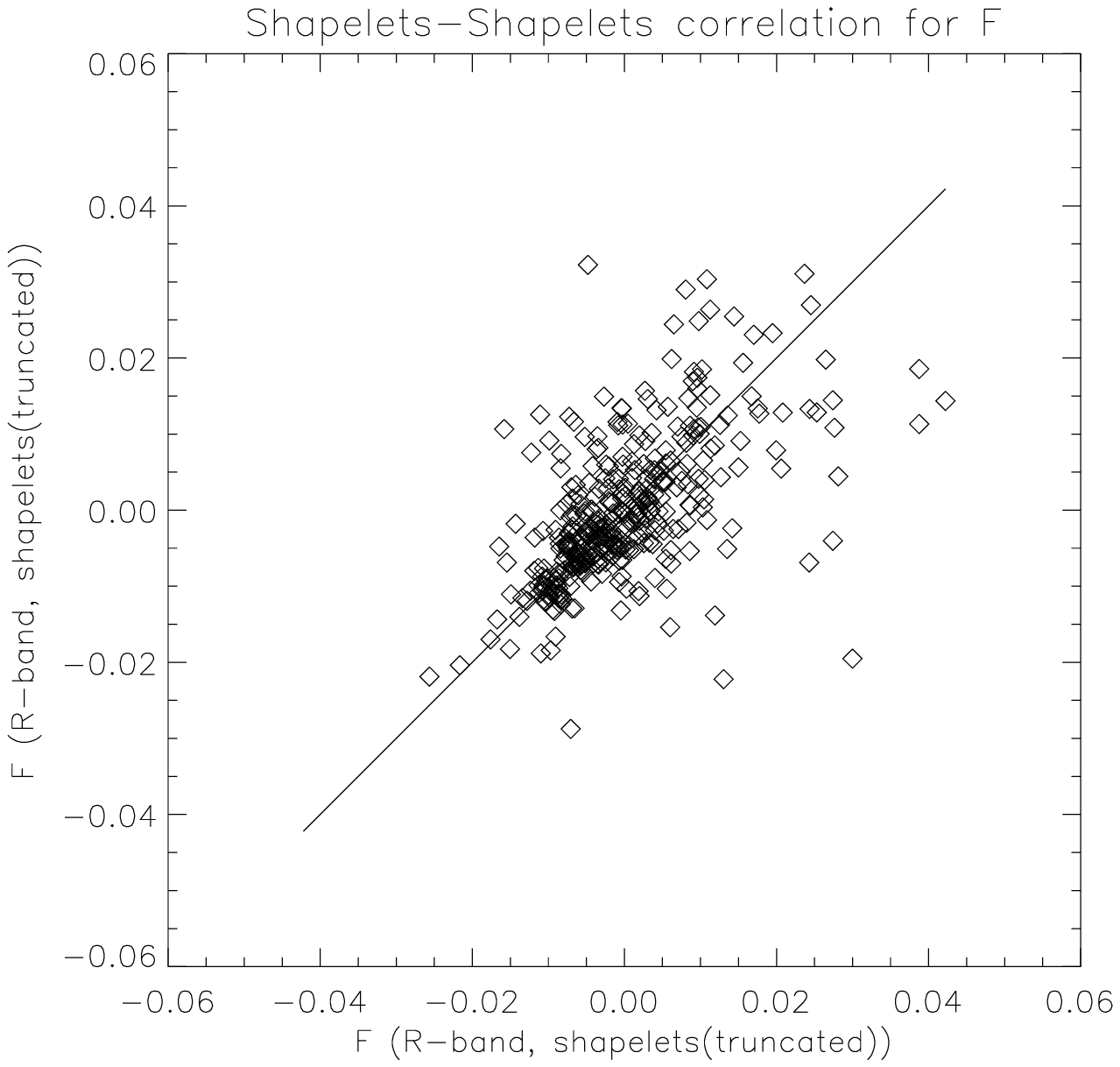,angle=0,height=2.5in}
  \psfig{figure=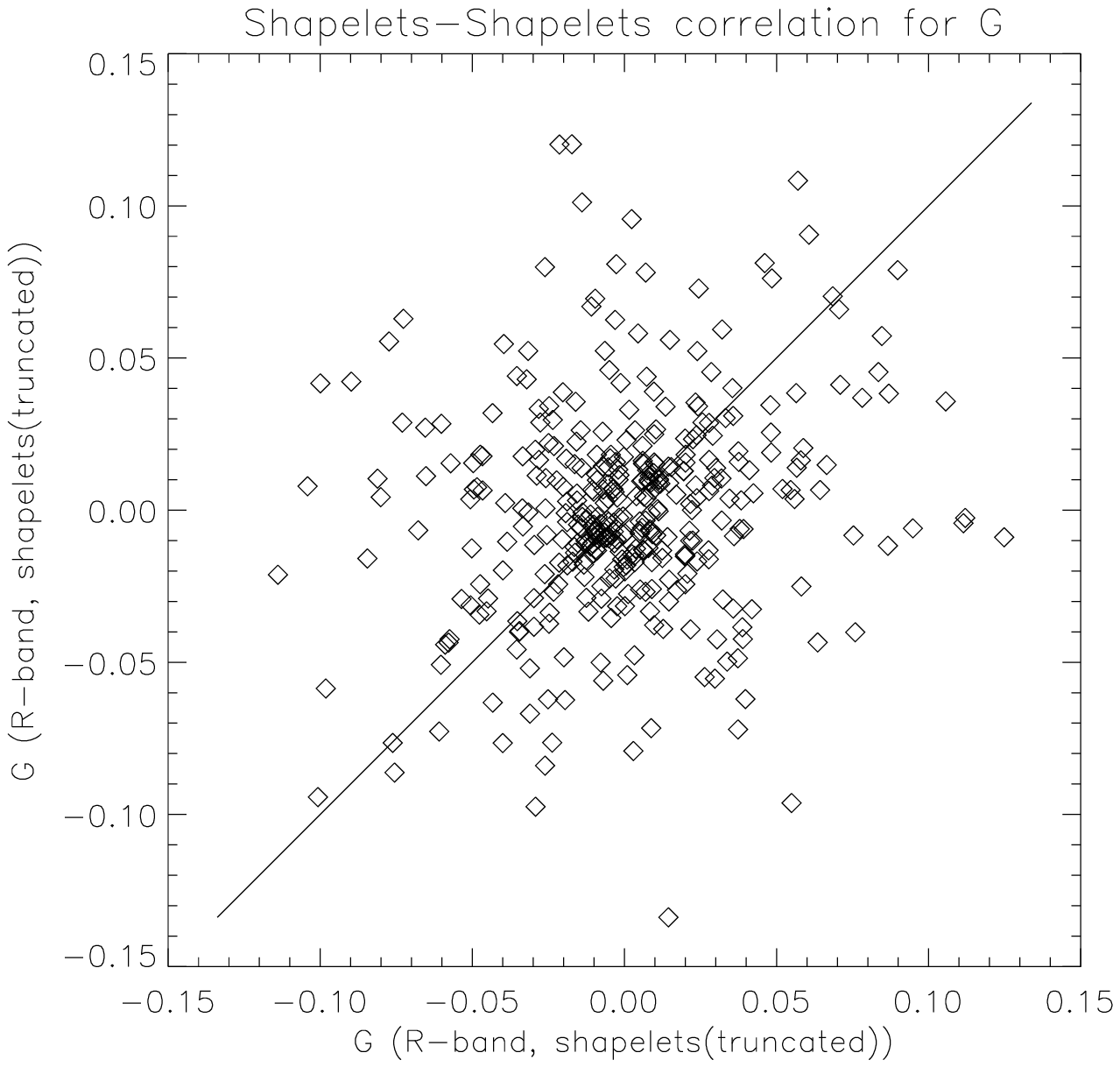,angle=0,height=2.5in}}
\caption{Truncated shapelets measurements of ${\cal F}$ and ${\cal G}$
  in the split images}
\label{split_shapelets}
\end{figure}

Figure~ \ref{split_moments} shows a comparison of the HOLICs
measurements made on each of the split images. These measurements are
well correlated: the Pearson correlation coefficient here is 0.37 for
$\cal{F}$ and 0.23 for $\cal{G}$. In Figure~ \ref{split_shapelets}, we
see a comparison of the shapelets measurements in these images, which
appear to be more strongly correlated (particularly for
$\cal{F}$). The Pearson correlation coefficients here are 0.58 for
$\cal{F}$ and 0.18 for $\cal{G}$.

\begin{figure}
\centerline{\psfig{figure=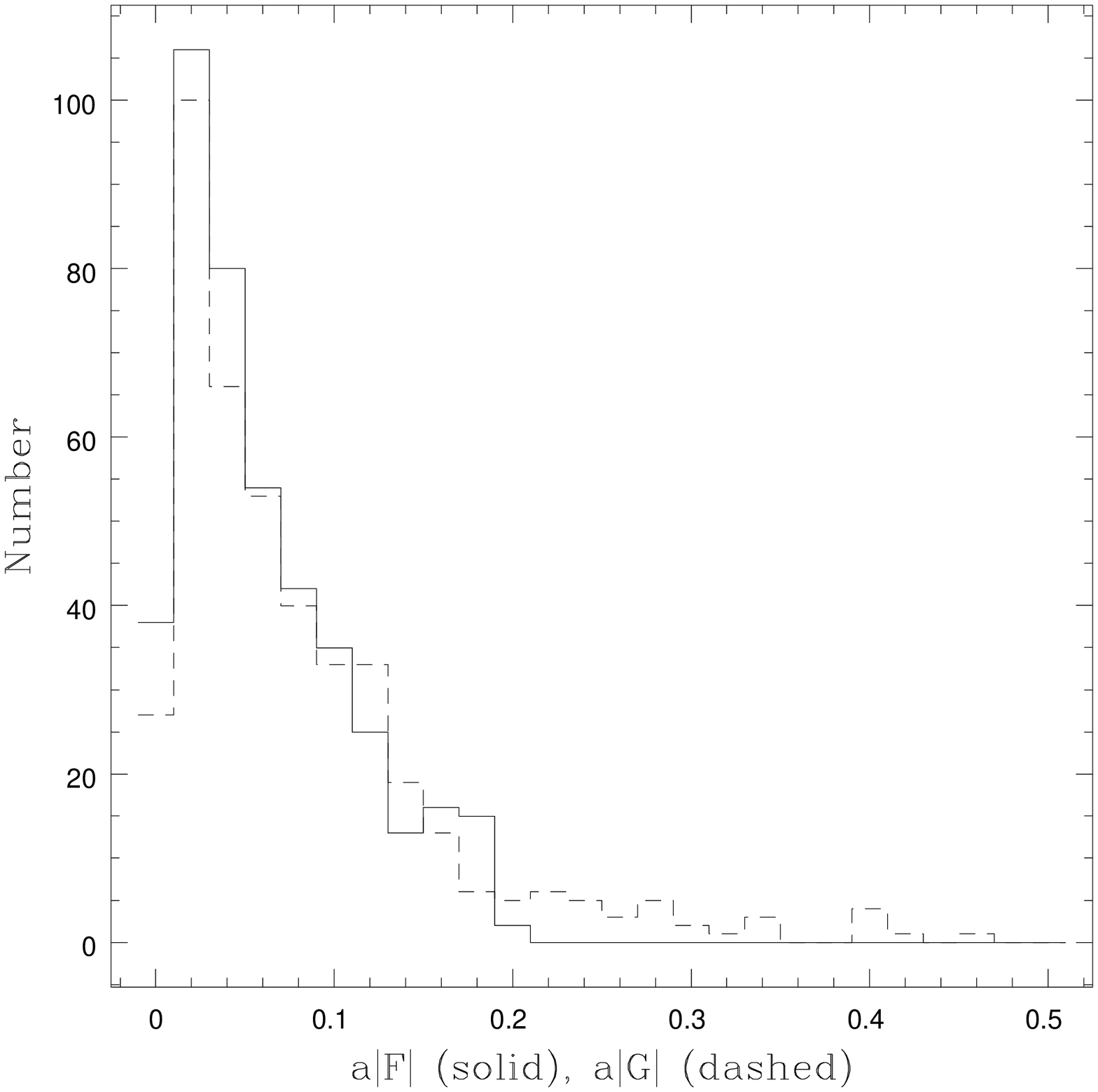,angle=0,height=3in}}
\caption{The histogram of $a|{\cal F}|$ and $a|{\cal G}|$ for our
  measured sample.  This represents an estimate of the intrinsic
  distribution of flexion among our source population.}
\label{fg:fhist}
\end{figure}

As motivated above, most of the ``noise'' in our measurements comes
from the intrinsic distribution of flexion within our sample.  Indeed,
using the HOLICs approach, we find:
\begin{equation}
\sigma_{a|F|+Noise}=0.05
\end{equation}
\begin{equation}
\sigma_{a|G|+Noise}=0.08
\end{equation}
The distribution function may be seen in Fig.~\ref{fg:fhist}.  Note
that this result includes noise.  However, we may estimate the
relative effect of photon noise on this scatter by using correlation
between frames.  That is:
\begin{equation}
\sigma_{a|F|}=\sqrt{\rho}\sigma_{a|F|+Noise}
\end{equation}
And thus, we find that our best estimate of the intrinsic scatter in
first flexion is:
\begin{equation}
\sigma_{a|F|}=0.03
\end{equation}
(as found in Goldberg \& Bacon 2005), and
\begin{equation}
\sigma_{a|G|}=0.04\ .
\end{equation}

The combination $a|{\cal F}|$ represents a dimensionless term, and
thus is independent of distance. 

It should also be noted that since these measurements are taken within
a cluster, the signal is included as well, and one might question
whether it is reasonable to estimate the intrinsic variability of
flexion from lensed images.  The intrinsic scatter in flexion was
originally measured in Goldberg \& Bacon (2005), and we merely confirm
the result here.  However, this is a reasonable thing to do, as
flexion drops off much more quickly than shear, and thus, even within
a rich cluster, the flexion signal is dominated by individual
galaxies.  Even at a separation of $1''$, the flexion from even a very
massive $300 km/s$ galaxy on an $a=0.4''$ source is about 0.05,
approximately the level of the intrinsic flexion.  Such separations
are relatively rare, however.

\section{Discussion}

\label{sec:discuss}

We have endeavored to present a detailed guide to measuring flexion in
real observations, with a focus on space-based imaging.  In the
process, we have taken a look at two different approaches to measuring
flexion: shapelets and HOLICs, with an eye toward which approach is
``better.''  From an idealized perspective of maximal signal-noise,
the answer is simple.  Shapelets produces a mode-by-mode comparison
which optimally averages to produce a unique estimate of flexion.
However, this result is complicated somewhat in two limits: blending,
which affects larger objects, and PSF convolution, which affects
smaller ones.

When images are blended, it is clear that we benefit by giving extra
weighting to those pixels near the center of the the object.  In that
sense, HOLICs can be said to produce more robust results.  Likewise,
despite an explicit PSF deconvolution algorithm, applying the
flexion inversion using shapelets results in inclusion of small scale
power which has been blended away through the atmosphere or
instrument.  We have discussed, above, how this might be alleviated by
only using relatively low order modes from the reconstruction in the
estimate of flexion.  However, doing so comes at the expense of some
(but by no means all), of the signal-noise advantage from shapelets.
Indeed, even using a relatively truncated form of the shapelets
analysis still produced greater correlation between independent images
of the same objects, and thus, cleaner estimates of the flexion.

However, one complication in the shapelets analysis is producing a
good shapelets decomposition in the first place.  While R. Massey's
shapelet code comes with an optimization routine to find the best fit
scaling parameter, $\beta$, the shapelet decomposition runs several
orders of magnitude slower than HOLICs.  For very large lensing
fields, this may prove a significant limitation, and thus, HOLICs
provides a fast, physically motivated, reasonably reliable
alternative.

\acknowledgements

This work was supported by NASA ATP NNG05GF61G and HST Archival grant
10658.01-A.  The authors would like to gratefully acknowledge useful
conversations with Jason Haaga, David Bacon \& Sanghamitra Deb, and
thank Richard Massey for thoughtful comments and the use of his
shapelets code.  We would also like to thank the anonymous referee
whose comments greatly improved the final draft.

\appendix

\section{Expanded Coefficients for HOLICs Analysis of Flexion}

In equation~(\ref{eq:Fsolve}), we state that the flexion may be solved
via inversion of the relation:
\begin{equation}
y={\cal M}x
\end{equation}
where $x$ is a vector of the desired flexion estimators, and $y$ is
the measure of the 3rd order HOLICs.  Here, we show the explicit form
of {\cal M}.

\begin{eqnarray}
M_{11}&=&\frac{1}{4}(9+8\eta_1)-\frac{33Q_{11}^2+14Q_{11}Q_{22}+Q_{22}^2+20Q_{12}^2}{4\xi}
\nonumber \\
M_{12}&=&2\eta_2-\frac{32Q_{12}Q_{22}+32Q_{11}Q_{12}}{4\xi}
\nonumber \\
M_{13}&=&\frac{1}{4}(2\eta_1+\lambda_1)-\frac{3Q_{11}^2-2Q_{11}Q_{22}-Q_{22}^2-4Q_{12}^2}{4\xi}
\nonumber \\
M_{14}&=&\frac{1}{4}(2\eta_2+\lambda_2)-\frac{2Q_{11}Q_{12}}{\xi}
\nonumber \\
M_{21}&=&2\eta_2-\frac{32Q_{12}Q_{22}+32Q_{11}Q_{12}}{4\xi}
\nonumber \\
M_{22}&=&\frac{1}{4}(-8\eta_1+9)-\frac{Q_{11}^2+14Q_{11}Q_{22}+20Q_{12}^2+33Q_{22}^2}{4\xi}
\nonumber \\
M_{23}&=&\frac{1}{4}(-2\eta_2+\lambda_2)-\frac{-2Q_{12}Q_{22}}{/\xi}
\nonumber \\
M_{24}&=&\frac{1}{4}(2\eta_1-\lambda_1)-\frac{(Q_{11}^2+4Q_{12}^2+Q_{11}Q_{22}-3Q_{22}^2)}{4\xi}
\nonumber \\
M_{31}&=&\frac{1}{4}(10\eta_1+7\lambda_1)-\frac{3(11Q_{11}^2-10Q_{11}Q_{22}-Q_{22}^2-20Q_{12}^2)}{4\xi}
\nonumber \\
M_{32}&=&\frac{1}{4}(-10\eta_2+7\lambda_2)-\frac{3(8Q_{11}Q_{12}-32Q_{12}Q_{22})}{4\xi}
\nonumber \\
M_{33}&=&\frac{3}{4}-\frac{3(-2Q_{11}Q_{22}+Q_{11}^2+Q_{22}^2+4Q_{12}^2)}{4\xi}
\nonumber \\
M_{34}&=&0
\nonumber \\
M_{41}&=&\frac{1}{4}(10\eta_2+7\lambda_2)-\frac{3(32Q_{11}Q_{12}-8Q_{12}Q_{22})}{4\xi}
\nonumber \\
M_{42}&=&\frac{1}{4}(10\eta_1-7\lambda_1)-\frac{3(Q_{11}^2+20Q_{12}^2+10Q_{11}Q_{22}-11Q_{22}^2)}{4\xi}
\nonumber \\
M_{43}&=&0
\nonumber \\
M_{44}&=&\frac{3}{4}-\frac{3(-2Q_{11}Q_{22}+Q_{11}^2+Q_{22}^2+4Q_{12}^2)}{4\xi}
\end{eqnarray}
where, as defined in Okura et al. (2006), we use:
\begin{equation}
\eta\equiv \frac{(Q_{1111}-Q_{2222})+2i(Q_{1112}+Q_{1222})}{\xi}
\end{equation}
\begin{equation}
\lambda\equiv\frac{(Q_{1111}-6Q_{1122}+Q_{2222})+4i(Q_{1112}-Q_{1222})}{\xi}
\end{equation}
Note that $\eta=0$ and $\lambda=0$ for all circularly symmetric distributions and even
those with no ellipticity but with flexion.

If we apply a Gaussian weighting with width, $\sigma_W$, to our moment
measurements, then ${\cal M}$ should be computed using the weighted
moments.  In addition, the following terms must be added:
\begin{eqnarray}
\Delta
  M_{11}&=&\frac{-3Q_{111111}-6Q_{111122}-3Q_{112222}+3Q_{22}Q_{1122}+9Q_{11}Q_{1111}+
  6Q12Q_{1112}+9Q_{11}Q_{1122}+6Q_{12}
  Q_{1222}-3Q_{22}Q_{1111}}{4\xi\sigma_W^2}\nonumber \\
\Delta
  M_{12}&=&\frac{-3Q_{111112}-6Q_{111222}-3Q_{122222}+3Q_{22}Q_{1112}+
  9Q_{11}Q_{1222}+3Q_{22}Q_{1222}+
  6Q_{12}Q_{1122}+
  9Q_{11}Q_{1112}+6Q_{12}Q_{2222}}
{4\xi\sigma_W^2}\nonumber \\
\Delta
  M_{13}&=&\frac{-Q_{111111}+2Q_{111122}+3Q_{112222}-3Q_{22}Q_{1122}+3Q_{11}Q_{1111}+
  2Q_{12}Q_{1112}-9Q_{11}Q_{1122}-6Q_{12}Q_{1222}+Q_{22}Q_{1111}}
{4\xi\sigma_W^2}\nonumber \\
\Delta
  M_{14}&=&\frac{-3Q_{111112}-2Q_{111222}+Q_{122222}+6Q_{12}Q_{1122}+9Q_{11}Q_{1112}-3Q_{11}Q_{1222}+3Q_{22}Q_{1112}-
    Q_{22}Q_{1222}-2Q_{12}Q_{2222}}
{4\xi\sigma_W^2} \nonumber \\
\Delta
  M_{21}&=&\frac{-3Q_{111112}-6Q_{111222}-3Q_{122222}+6Q_{12}Q_{1122}+3Q_{11}Q_{1112}+9Q_{22}Q_{1112}+3Q_{11}Q_{1222}+
    9Q_{22}Q_{1222}+6Q_{12}Q_{1111}}
{4\xi\sigma_W^2} \nonumber \\
\Delta
  M_{22}&=&\frac{-3Q_{111122}-6Q_{112222}-3Q_{222222}+6Q_{12}Q_{1112}+3Q_{11}Q_{2222}+6Q_{12}Q_{1222}+9Q_{22}Q_{1122}+
    3Q_{11}Q_{1122}+9Q_{22}Q_{2222}}
{4\xi\sigma_W^2} \nonumber \\
\Delta
  M_{23}&=&\frac{-Q_{111112}+2Q_{111222}+6Q_{122222}-6Q_{12}Q_{1122}+Q_{11}Q_{1112}+3Q_{22}Q_{1112}-3Q_{11}Q_{1222}-
9Q_{22}Q_{1222}+2Q_{12}Q_{1111}}
{4\xi\sigma_W^2} \nonumber \\
\Delta
  M_{24}&=&\frac{-3Q_{111122}-2Q_{112222}+Q_{222222}+9Q_{22}Q_{1122}+3Q_{11}Q_{1122}-Q_{11}Q_{2222}+6Q_{12}Q_{1112}-
    2Q_{12}Q_{1222}-3Q_{22}Q_{2222}}
{4\xi\sigma_W^2} \nonumber \\
\Delta
  M_{31}&=&\frac{-3Q_{111111}+6Q_{111122}+9Q_{112222}-9Q_{22}Q_{1122}+9Q_{11}Q_{1111}-18Q_{12}Q_{1112}+9Q_{11}Q_{1122}-
18Q_{12}Q_{1222}-9Q_{22}Q_{1111}}
{4\xi\sigma_W^2} \nonumber \\
\Delta
  M_{32}&=&\frac{-3Q_{111112}+6Q_{111222}+9Q_{122222}-9Q_{22}Q_{1112}+9Q_{11}Q_{1222}-9Q_{22}Q_{1222}-18Q_{12}Q_{1122}+
    9Q_{11}Q_{1112}-18Q_{12}Q_{2222}}
{4\xi\sigma_W^2} \nonumber \\
\Delta
  M_{33}&=&\frac{-Q_{111111}+6Q_{111122}-9Q_{112222}+9Q_{22}Q_{1122}+3Q_{11}Q_{1111}-6Q_{12}Q_{1112}-9Q_{11}Q_{1122}+
    18Q_{12}Q_{1222}-3Q_{22}Q_{1111}}
{4\xi\sigma_W^2} \nonumber \\
\Delta
  M_{34}&=&\frac{-3Q_{111112}+10Q_{111222}-3Q_{122222}-18Q_{12}Q_{1122}+9Q_{11}Q_{1112}-3Q_{11}Q_{1222}-9Q_{22}Q_{1112}+
    3Q_{22}Q_{1222}+6Q_{12}Q_{2222}}
{4\xi\sigma_W^2} \nonumber \\
\Delta
  M_{41}&=&\frac{-9Q_{111112}-6Q_{111222}+3Q_{122222}+18Q_{12}Q_{1122}+9Q_{11}Q_{1112}-9Q_{22}Q_{1112}+9Q_{11}Q_{1222}-
    9Q_{22}Q_{1222}+18Q_{12}Q_{1111}}
{4\xi\sigma_W^2} \nonumber \\
\Delta
  M_{42}&=&\frac{-9Q_{111122}-6Q_{112222}+3Q_{222222}+18Q_{12}Q_{1112}+9Q_{11}Q_{2222}+18Q_{12}Q_{1222}-9Q_{22}Q_{1122}+
9Q_{11}Q_{1122}-9Q_{22}Q_{2222}}
{4\xi\sigma_W^2} \nonumber \\
\Delta
  M_{43}&=&\frac{-3Q_{111112}+10Q_{111222}-3Q_{122222}+18Q_{12}Q_{1122}+3Q_{11}Q_{1112}-3Q_{22}Q_{1112}-9Q_{11}Q_{1222}+
9Q_{22}Q_{1222}+6Q_{12}Q_{1111}}
{4\xi\sigma_W^2} \nonumber \\
\Delta
  M_{44}&=&\frac{-9Q_{111122}+6Q_{112222}-Q_{222222}+9Q_{22}Q_{1122}+9Q_{11}Q_{1122}-3Q_{11}Q_{2222}+18Q_{12}Q_{1112}-
6Q_{12}Q_{1222}+3Q_{22}Q_{2222}}{4\xi\sigma_W^2} 
\end{eqnarray}

\end{document}